\documentclass[11pt]{article}

\usepackage[a4paper, margin=1in]{geometry}
\usepackage{amsmath, amssymb, amsfonts, amsthm}
\usepackage{mathtools}
\usepackage{booktabs}
\usepackage{longtable}   
\usepackage{multirow}
\usepackage{array}
\usepackage{graphicx}
\usepackage{xcolor}
\usepackage[hidelinks]{hyperref}
\usepackage{caption}
\usepackage{subcaption}
\usepackage{enumitem}
\usepackage{microtype}
\usepackage{authblk}
\usepackage{cite}
\usepackage{float}
\usepackage[normalem]{ulem}

\definecolor{forestgreen}{rgb}{0.33,0.61,0.34}

\title{Baseline-referenced spatial early warning signals for tipping points on heterogeneous networks}
\author[1]{Tharusha Bandara}
\author[2]{Shilong Yu}
\author[1,2,3,*]{Naoki Masuda}
\affil[1]{Gilbert S.\ Omenn Department of Computational Medicine and Bioinformatics, University of Michigan, Ann Arbor, MI 48109, USA}
\affil[2]{Department of Mathematics, University of Michigan, Ann Arbor, MI 48109, USA}
\affil[3]{Center for Computational Social Science, Kobe University, Kobe, Hyogo 657-8501, Japan}
\affil[*]{Corresponding author. E-mail: \texttt{naokimas@umich.edu}}
\date{}

\begin{document}

\maketitle

\begin{abstract}
Anticipating tipping points in complex systems is difficult because many early warning signals require long time series, which are often unavailable in practice. Spatial early warning signals offer an alternative by using a single snapshot across many interacting elements, or nodes. However, their performance in heterogeneous systems is often inconsistent because raw node states reflect both dynamical changes associated with an approaching transition and static heterogeneity induced by network structure. Here, we propose a baseline-referenced framework for spatial early warning signals. The method compares each node's state with its own baseline far from the tipping point before computing a spatial statistic, thus reducing network-structure-induced variation. We evaluate baseline-referenced variants of five classical spatial early warning signals across diverse tipping scenarios and networks, and find that baseline referencing markedly improves variance-based spatial signals. The best variants increase consistently and progressively toward tipping points across different scenarios, outperform a single-node temporal variance that requires long time series, and retain high performance even when up to 80\% of nodes are omitted from observation. These results provide a practical route for using spatial early warning signals in heterogeneous networked systems when dense temporal monitoring or complete network-wide observation is infeasible, as is often the case in real applications.
\end{abstract}


\section*{Significance statement}

Sudden transitions in ecosystems, epidemics, climate subsystems, and other complex systems can be difficult to anticipate because standard early warning signals often require long time series. Spatial early warning signals reduce this burden by using measurements across many locations or interacting components at a single time point, but previous research has shown that they often fail in heterogeneous networks. We show that comparing each component with its own baseline before computing a spatial signal makes variance-based early warning signals substantially more reliable across diverse dynamical systems and networks. We also show that monitoring only a modest fraction of components can retain most of the warning performance. This approach may help make early warning analysis more feasible in data-limited real systems.

\section{Introduction}\label{introduction}

Tipping points occur when a complex dynamical system enters a qualitatively different state. Such transitions are believed to underlie sudden regime shifts in natural and societal systems and should therefore be heeded. For example, the sudden shift of a lake from a clear-water state to a turbid, algae-dominated state once the nutrient load crosses a critical threshold is a classic ecological tipping point \cite{Scheffer2001Nature}. The Greenland Ice Sheet is considered a climate tipping element: beyond a warming threshold, melt-elevation feedback can drive self-reinforcing mass loss, committing the system to substantial and potentially irreversible sea-level rise \cite{Lenton2008PNAS}. The potentially severe consequences of such abrupt transitions have motivated numerous approaches for identifying them before they occur, collectively referred to as early warning signals (EWSs) \cite{Scheffer2009Nature, dakos2012methods, dakos2015resilience}, despite important caveats \cite{boettiger2013early, kefi2013early, rietkerk2025ambiguity}.

Many EWSs are based on critical slowing down: as a dynamical system approaches a bifurcation, its recovery from small perturbations becomes slower. Classical temporal EWSs, such as variance and lagged autocorrelation, exploit this phenomenon by estimating statistics from time series observed under slowly changing environmental conditions \cite{Scheffer2009Nature, dakos2012methods, Dakos2012TheorEcol}. Related network-based~\cite{Patterson2021AmNat, Aparicio2021PNAS, MacLaren2023JRoySocInterface, masuda2024anticipating}, model-fitting \cite{boettiger2012quantifying, hessler2022bayesian}, and machine-learning \cite{kong2021machine, patel2023using, liu2024early} approaches have expanded the EWS toolkit, but many of them still require relatively long time series. Even estimating a single sample variance reliably may require many samples \cite{JCGM100_2008}, which is often impractical when observations are expensive, destructive, or slow to obtain \cite{Biggs2009PNAS, Dakos2010TheorEcol}. Rolling windows are commonly used to obtain temporal EWSs across environmental conditions \cite{dakos2012methods, buelo2022evaluating, gsell2016evaluating}, but they also smooth the signal and can reduce sensitivity to an approaching transition \cite{curtiss2023rising}.

Spatial EWSs offer a complementary route by replacing repeated sampling over time with sampling across space or across interacting elements of a system. In each environmental condition, one observes many distinct elements, which we call nodes, and computes a spatial statistic from a single snapshot \cite{Dakos2010TheorEcol, dai2013slower, kefi2014early, nijp2019spatial}. For example, the spatial variance of node states is a direct spatial analogue of temporal variance and requires only one sample per node. Most theoretical and simulation studies of spatial EWSs, however, have considered ecological dynamics on regular lattices or similar homogeneous spatial domains \cite{Dakos2010TheorEcol, kefi2014early, fernandez2009catastrophic, storch2022topological}. Empirical systems are rarely so regular: ecological, epidemiological, biological, infrastructural, and social systems are often organized as heterogeneous networks \cite{Newman2018book}. Recent numerical studies have shown that classical spatial EWSs can behave inconsistently on such networks, with no universally reliable signal across dynamical systems, networks, control parameters, and tipping scenarios, even if tipping occurs as a conventional bifurcation of dynamical systems \cite{maclaren2025applicability, robinson2025assessing}.
%
%
Thus, the main question is no longer only which existing spatial EWS performs best, but how spatial EWSs can be made robust to network heterogeneity.

Here, we develop and test a simple strategy for making spatial EWSs more reliable on heterogeneous networks. Our key idea is to reference each node to its own baseline far from the tipping point before computing a spatial statistic, thus reducing the contribution of static, network-structure-induced heterogeneity in node states. We construct baseline-referenced variants of classical spatial EWSs and evaluate them across stochastic dynamical-system models, networks, and tipping scenarios. We find that a baseline-referenced spatial variance is robust across these settings and can be computed from a modest fraction of nodes with little loss of early-warning performance. These results provide a practical prescription for spatial EWSs on heterogeneous networks and substantially extend previous benchmarking studies by turning network heterogeneity from a source of unreliability into a feature that can be explicitly controlled.

\section{Results}\label{sec:newresults}

\subsection{Classical spatial early warning signals are unreliable across networks}\label{sub:existing_fail}

Before presenting our method, we briefly illustrate why classical spatial EWSs are insufficient on general networks; also see our previous study \cite{maclaren2025applicability}. We use the stochastic coupled double-well dynamics on networks \cite{MacLaren2023JRoySocInterface, masuda2024anticipating, maclaren2025applicability},
\begin{equation}\label{eq:doublewell}
dx_i = \left[ -(x_i - r_1)(x_i - r_2)(x_i - r_3) + D \sum_{j=1}^N A_{ij} x_j + u \right] dt + \sigma\, dW_i,
\end{equation}
where $x_i$ is the state of the $i$th node, $A=(A_{ij})$ is the adjacency matrix of the network, $D$ is the coupling strength, $u$ is a stress parameter, $W_i$ is an independent standard Wiener process, and $\sigma$ is the noise strength. We assume $r_1 < r_2 < r_3$, so that, in the absence of coupling, stress, and noise, each node has a lower stable equilibrium at $r_1$ and an upper stable equilibrium at $r_3$. The coupled double-well dynamics has been used to model collective social dynamics \cite{brummitt2015coupled} and networks of climate subsystems \cite{wunderling2022recurrent}. We set $r_1=1$, $r_2=3$, $r_3=5$, and $\sigma=0.1$. For demonstration, we drive the system toward a tipping point in two ways: by gradually decreasing the coupling $D$ with $u=-5$, or by gradually decreasing the stress $u$ with $D=0.05$. In both cases, we initialize each simulation at the upper equilibrium, i.e., $x_i=5$ for all $i$, and record the equilibrium node states $x_i^*$ over the home range, defined as the range of the control parameter over which all nodes remain in the upper state, up to the first tipping event.

\begin{figure}[htbp]
    \centering
    \includegraphics[width=\textwidth]{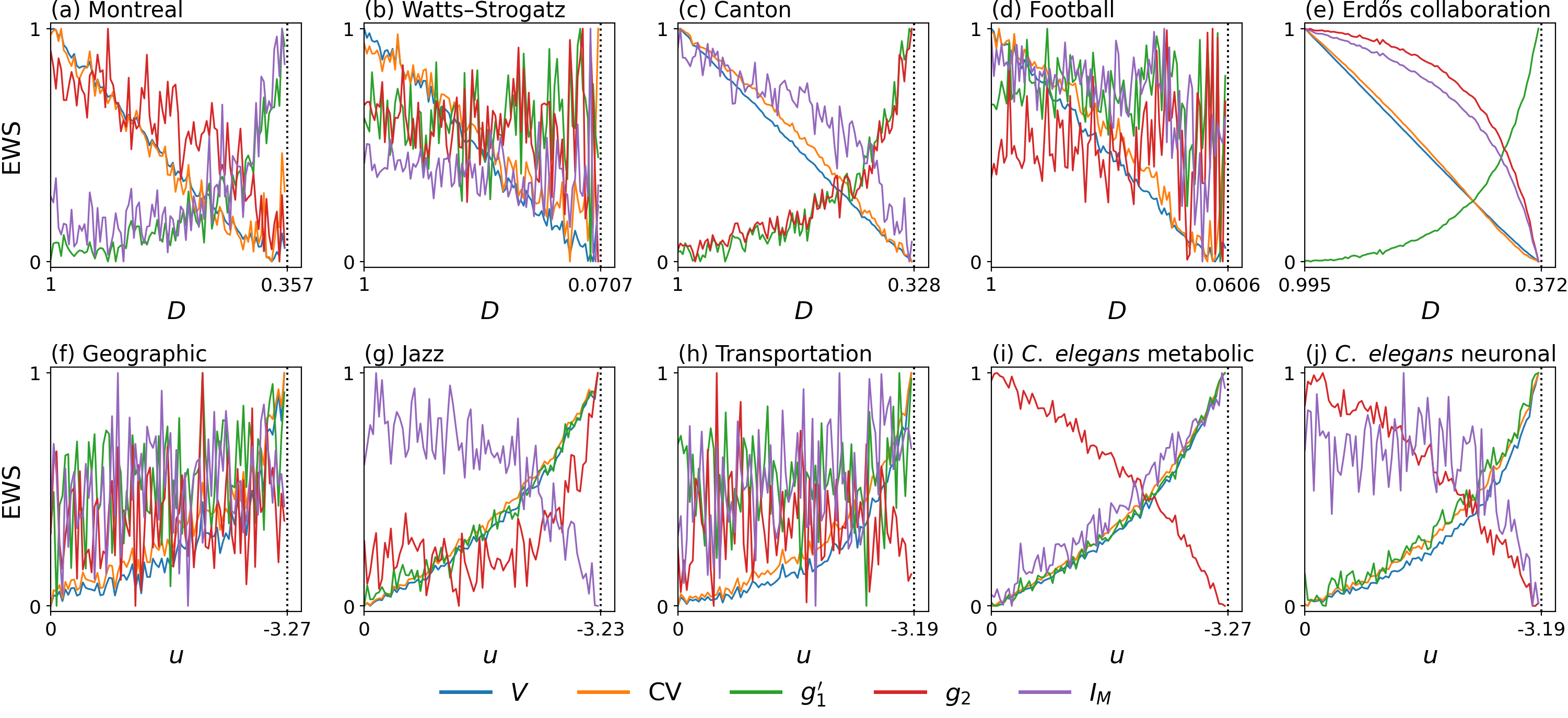}
\caption{Inconsistent behavior of classical spatial EWSs across networks. We use the coupled double-well dynamics. Each panel shows five EWSs over a range of the control parameter, i.e., the home range, with the control parameter running from far from the tipping point (left) to the first tipping point (right; dotted line). Each curve is min--max normalized to $[0,1]$ in each panel. (a)--(e) Descending coupling strength $D$ for the Montreal ($N=29$), Watts--Strogatz ($N=100$), Canton ($N=109$), football ($N=115$), and Erd\H{o}s collaboration ($N=6927$) networks. (f)--(j) Descending stress $u$ for the geographic ($N=49$), jazz player ($N=198$), transportation ($N=369$), \textit{C. elegans} metabolic ($N=453$), and \textit{C. elegans} neuronal ($N=460$) networks.}
    \label{fig:existing}
\end{figure}

For each network, we computed five widely used spatial EWSs, i.e., the spatial variance $V$ (or spatial standard deviation $\sqrt{V}$) \cite{guttal2009spatial, buelo2018modeling, eby2017alternative, litzow2017indications}, coefficient of variation (CV) \cite{dai2013slower, litzow2008increased, rindi2018experimental}, sign-adjusted skewness $g_1'$ \cite{guttal2009spatial, buelo2018modeling}, kurtosis $g_2$ \cite{litzow2017indications, buelo2018modeling}, and Moran's $I$, denoted by $I_M$ \cite{legendre1989spatial, Dakos2010TheorEcol, okabe2015spatial}, across the home range. A useful EWS should increase steadily and progressively as the dynamical system approaches the tipping point. Figure~\ref{fig:existing} shows these five EWSs for ten representative networks: five under descending $D$ (Fig.~\ref{fig:existing}(a)--(e)) and five under descending $u$ (Fig.~\ref{fig:existing}(f)--(j)).

The behavior of the five EWSs is strikingly inconsistent across networks. Under descending $D$ (Fig.~\ref{fig:existing}(a)--(e)), the spatial variance and CV systematically decrease toward the tipping point instead of increasing, i.e., they fail outright, because reducing $D$ shrinks the coupling-induced spread of the node states. The EWSs $g_1'$, $g_2$, and $I_M$ either increase or decrease as $D$ decreases, depending on the network. Under descending $u$ (Fig.~\ref{fig:existing}(f)--(j)), the variance and CV increase as desired, yet $g_1'$, $g_2$, and $I_M$ remain unreliable, ranging from essentially flat (Fig.~\ref{fig:existing}(h)) to strongly increasing or decreasing, depending on the network. No single classical spatial EWS works consistently across networks and tipping scenarios; the best signal depends not only on the dynamical-system model, the control parameter, and its direction, but also on the network \cite{maclaren2025applicability, robinson2025assessing}.
%
%
This unreliability motivates the robust approach that we develop in the next section.

\subsection{Baseline-referenced spatial EWSs}\label{sub:baseline}

The failure of the classical spatial EWSs illustrated in Fig.~\ref{fig:existing} has a common root. A single snapshot of the node states, $\{x_1^*,\ldots,x_N^*\}$, is dominated by heterogeneity in the network structure. Mainly because of degree heterogeneity, the equilibrium node states differ systematically even far from any tipping point. This baseline heterogeneity varies from network to network. A spatial statistic computed directly on $\{x_i^*\}$ therefore mixes the effect of network structure with the genuine dynamical changes that precede a tipping point; see \cite{Clarke2026JPhysComplexity} for related discussion. For example, as the coupling strength $D$ decreases, the coupling-induced spread of $\{x_i^*\}$ shrinks. Therefore, the raw spatial variance, as well as the CV, decreases even as the dynamical system approaches its tipping point (blue and orange lines in Fig.~\ref{fig:existing}(a)--(e)). Note that decreasing inter-node coupling, at least for some node pairs, is an established route to population collapse in networks \cite{gao2016universal, lever2020foreseeing}, and connectivity is known to suppress fluctuation-based spatial EWSs \cite{dai2013slower}.

Therefore, to mitigate the contribution of network structure to spatial EWSs, we reference each node to its own baseline. We define the baseline of the $i$th node by
\begin{equation}\label{eq:baseline}
b_i = \frac{1}{\ell}\sum_{\ell'=1}^{\ell} x_i^*(c_{\ell'}),
\end{equation}
i.e., the mean of $x_i^*$ over the first $\ell$ control-parameter values, $c_1,\ldots,c_\ell$, of the home range, which are far from the tipping point. We use $\ell=5$ throughout. We then compute each spatial EWS not on $\{x_i^*\}$ but on the baseline-referenced node states, in one of two variants. The difference variant uses $x_i^*-b_i$, and the ratio variant uses $x_i^*/b_i$. Taking the spatial variance as an example, we replace the original variance $V=\mathrm{Var}_i(x_i^*)$ by
\begin{align}
\label{eq:vvariants}
V_\Delta &= \mathrm{Var}_i\!\left(x_i^*-b_i\right), \qquad \text{or}\\
V_{\rm rel} &= \mathrm{Var}_i\!\left(x_i^*/b_i\right),
\end{align}
where $\mathrm{Var}_i$ denotes the sample variance over the nodes. Both variants remove the static, network-specific component, $b_i$, additively in the case of $V_\Delta$ and multiplicatively in the case of $V_{\rm rel}$. The resulting spatial EWSs are expected to capture dynamical contributions to the broadening of the distribution of node states as the dynamical system approaches the tipping point.

\begin{figure}[htbp]
    \centering
    \includegraphics[width=\textwidth]{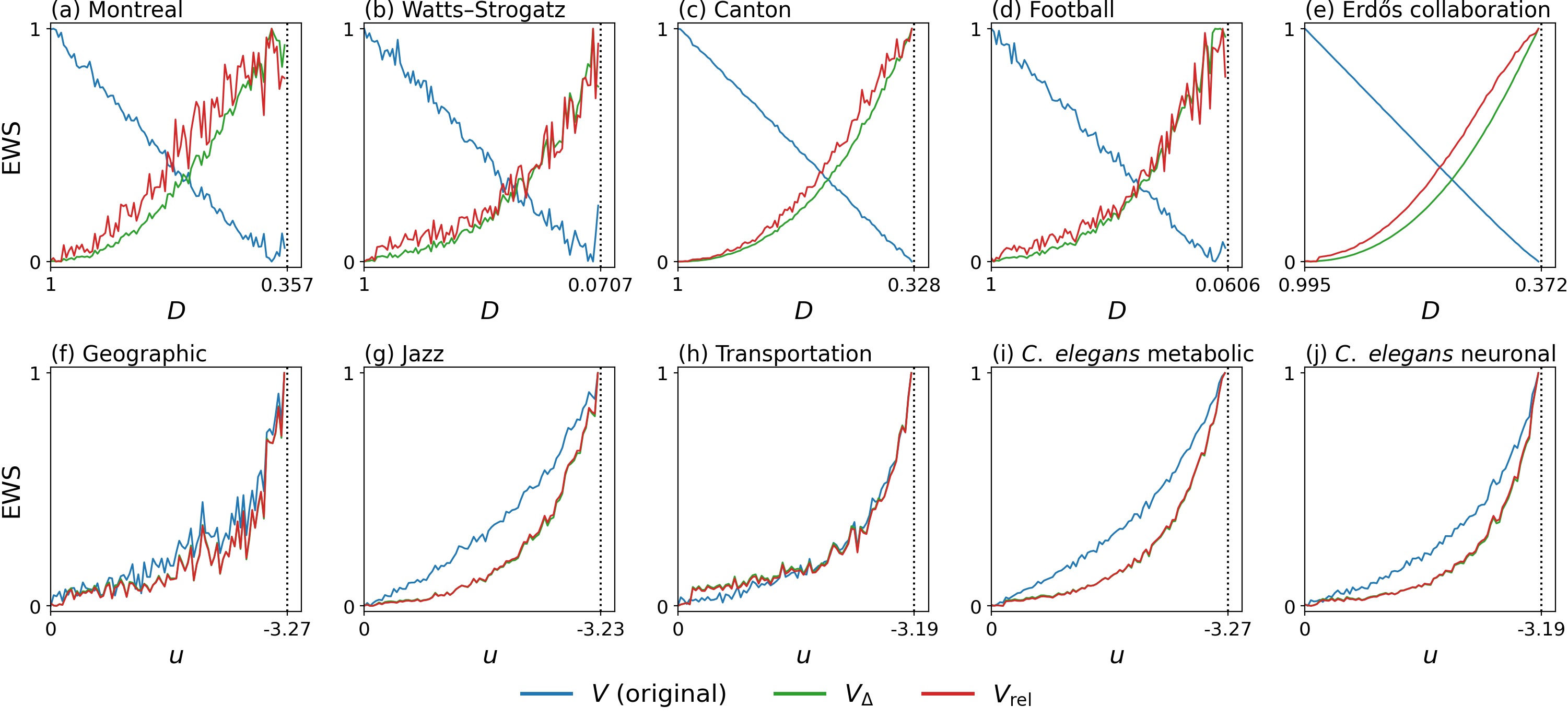}
\caption{Sample behavior of the baseline-referenced variants of the variance, $V_\Delta$ and $V_{\rm rel}$. We use the same ten pairs of descending control parameter and network as in Fig.~\ref{fig:existing}. (a)--(e) Descending coupling strength $D$. (f)--(j) Descending stress $u$. Each panel shows the original spatial variance $V$ (blue), its difference variant $V_\Delta$ (green), and its ratio variant $V_{\rm rel}$ (red), each min--max normalized to $[0,1]$. The blue curves are the same as those shown in Fig.~\ref{fig:existing}.}
    \label{fig:varvariants}
\end{figure}

Figure~\ref{fig:varvariants} shows the behavior of $V_{\Delta}$ and $V_{\rm rel}$ together with the original spatial variance, $V$, for the same ten combinations of control parameter and network as in Fig.~\ref{fig:existing}. The results for $V$ (shown in blue) are identical to those in Fig.~\ref{fig:existing}. Although $V$ decreases toward the tipping point under descending $D$ (Fig.~\ref{fig:varvariants}(a)--(e)), both $V_\Delta$ and $V_{\rm rel}$ increase steadily and progressively toward the tipping point in all five networks. Furthermore, $V_\Delta$ and $V_{\rm rel}$ also increase steadily under descending $u$ (Fig.~\ref{fig:varvariants}(f)--(j)). These results suggest that the baseline-referenced variances, $V_\Delta$ and $V_{\rm rel}$, are more consistent across networks and simulation conditions than the original spatial variance.

We now comprehensively evaluate the baseline-referenced EWSs across types of spatial EWSs, simulation conditions, and networks. A simulation condition is a combination of a dynamical-system model, a control parameter, and the direction in which the control parameter is varied. We use four stochastic dynamical systems on networks: the coupled double-well dynamics given by Eq.~\eqref{eq:doublewell}, a model of mutualistic species interactions, a susceptible--infectious--susceptible (SIS) model for epidemic spreading, and a gene-regulatory model. For the double-well, mutualistic, and gene-regulatory models, the control parameter is either the coupling strength $D$ or the stress $u$. For the SIS model, we use only $D$, which is equivalent to the infection rate, because a uniform external stress is not physically relevant for an epidemic. The direction is either ascending, in which the control parameter is gradually increased starting from a state in which all nodes are in their lower state, or descending, in which the control parameter is gradually decreased starting from an upper state. For the coupled double-well dynamics and the SIS model, both transitions are of interest, so we use both directions. For the mutualistic and gene-regulatory models, the transition of practical interest is the collapse of the established upper state, i.e., a loss of resilience such as a population collapse, so we consider only descending simulations \cite{gao2016universal, maclaren2025applicability}. These choices yield ten simulation conditions in total.

We define the difference and ratio variants in exactly the same manner for the other four classical spatial EWSs. In other words, for each EWS, we first replace every node state $x_i^*$ by $x_i^*-b_i$ or by $x_i^*/b_i$, and then compute the EWS, i.e., the CV, sign-adjusted skewness ($g_1'$), kurtosis ($g_2$), or Moran's $I$ ($I_{\text{M}}$), on the transformed node states.

\begin{figure}[htbp]
    \centering
    \includegraphics[width=\textwidth]{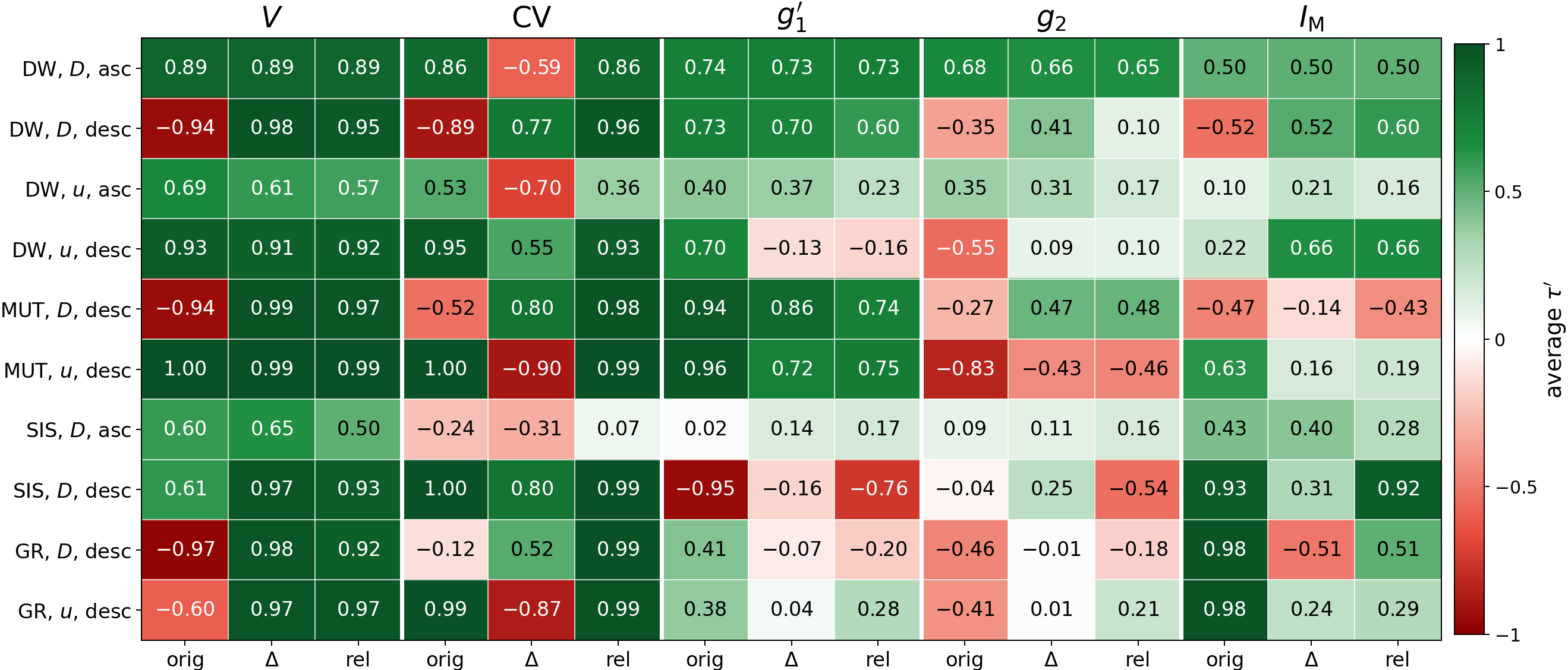}
\caption{Performance of the baseline-referenced spatial EWSs. Each cell shows the sign-adjusted Kendall's $\tau$ (i.e., $\tau'$) averaged over the 40 networks. The rows correspond to the ten simulation conditions (DW: double-well; MUT: mutualistic interaction; SIS: susceptible--infectious--susceptible; GR: gene-regulatory; $D$ or $u$: control parameter; asc/desc: ascending/descending). Each column corresponds to one of the five classical spatial EWSs in one of three variants: original (orig), difference ($\Delta$, computed on $x_i^*-b_i$), or ratio (rel, computed on $x_i^*/b_i$).}
    \label{fig:tauheat}
\end{figure}

We show in Fig.~\ref{fig:tauheat} the sign-adjusted Kendall's $\tau$, denoted by $\tau'$, as a performance measure, averaged over the 40 networks for each combination of EWS and simulation condition. (See section S1 for the networks.) A larger $\tau'$ value indicates a better EWS regardless of the simulation direction, with $\tau'\in[-1,1]$. Figure~\ref{fig:tauheat} indicates that baseline referencing can notably improve the two variance-based EWSs, i.e., $V$ and CV, but not the other three EWSs. All five original spatial EWSs and the baseline-referenced variants of the three non-variance-based EWSs, i.e., $g_1'$, $g_2$, and $I_{\text{M}}$, are unreliable. Specifically, whether these EWSs increase, decrease, or behave otherwise toward the tipping point strongly depends on the simulation condition, as indicated by opposite signs and a wide range of $\tau'$ values across simulation conditions. In contrast, $V_\Delta$, $V_{\rm rel}$, and $\text{CV}_{\rm rel}$ increase toward the tipping point across all simulation conditions, at least on average over the networks, although $\text{CV}_{\rm rel}$ has a nearly zero $\tau'$ value for one simulation condition. We obtain qualitatively the same results when we compute $\tau'$ over only the second half of the home range, i.e., the half closer to the tipping point (see Section~S2 for the results).

\subsection{Distinguishing acceleration from a linear trend}\label{sub:supf}

A large Kendall's $\tau'$ indicates that an EWS increases approximately monotonically toward the tipping point. However, it does not distinguish a linear increase from an accelerating, i.e., nonlinear, increase. This distinction matters for anticipation: a signal that merely rises linearly throughout the home range does not by itself indicate that a tipping point is imminent, whereas critical slowing down predicts that a true precursor should rise increasingly steeply as the tipping point is approached \cite{Scheffer2009Nature, Bury2020JRSocInterface, masuda2026tipmoc}. Therefore, for the best-performing EWSs, we ask whether they show a significant positive change in slope, i.e., a steepening, before the tipping point rather than merely a large $\tau'$.

We focus on the three best variants identified above, namely $V_\Delta$, $V_{\rm rel}$, and $\mathrm{CV}_{\rm rel}$. For comparison, we also include the original, i.e., non-baseline-referenced, spatial variance $V$ and CV. For each simulation condition and network, we applied the sup-$F$ test for a structural change at an unknown breakpoint \cite{quandt1960tests, andrews1993tests, zeileis2002strucchange} to the EWS over the home range. 
We regard the resulting sup-$F$-based procedure as successful when the test detects a statistically significant change in the regression coefficients ($p<0.05$) and, at the breakpoint maximizing the $F$ statistic, the fitted slope is non-negative before the breakpoint and larger after the breakpoint.

\begin{figure}[htbp]
    \centering
    \includegraphics[width=\textwidth]{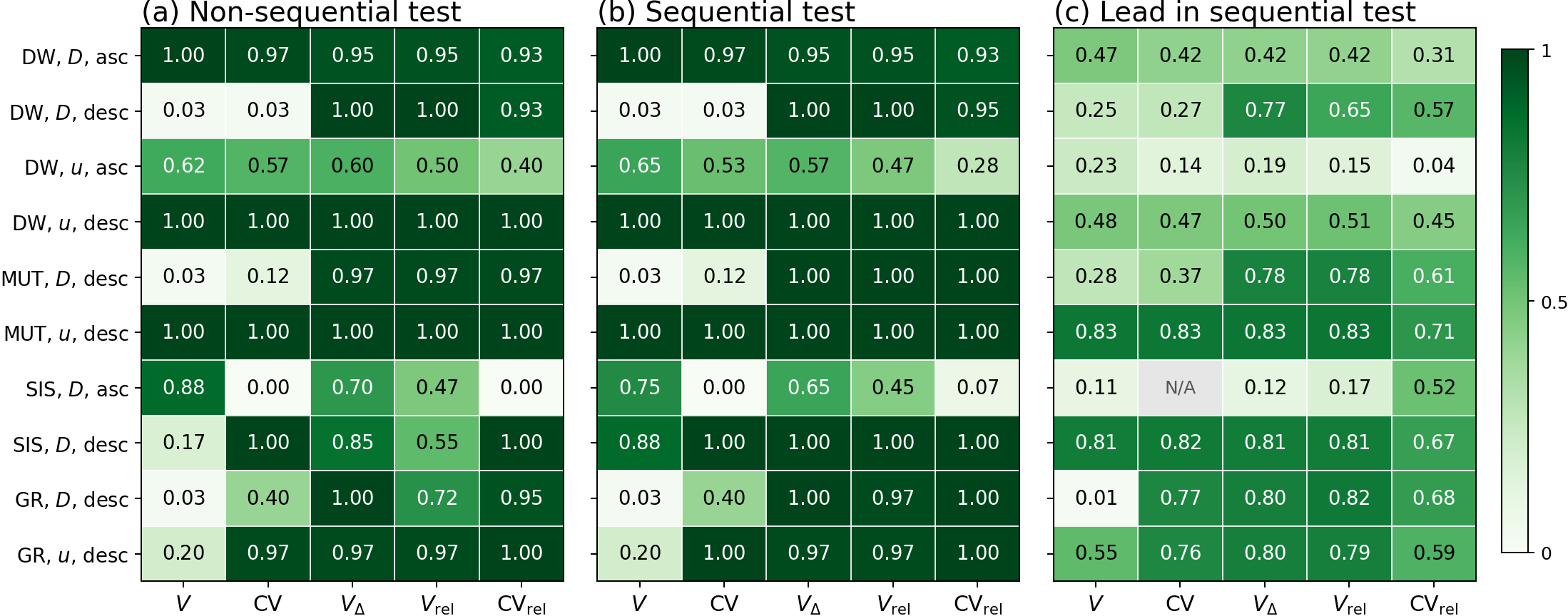}
\caption{Detection of a significant positive steepening of the EWS as the tipping point is approached, using the sup-$F$-based procedure, for the original $V$ and CV and the three baseline-referenced EWS variants, $V_\Delta$, $V_{\rm rel}$, and $\mathrm{CV}_{\rm rel}$. Rows correspond to the ten simulation conditions; abbreviations are as in Fig.~\ref{fig:tauheat}. (a) Fraction of the 40 networks for which the non-sequential test, using the entire home range, identifies a positive steepening. (b) Fraction of the networks for which the sequential test identifies a positive steepening before the tipping point. (c) Average lead time at the positive-steepening alarm in the sequential test, i.e., the fraction of the home range still remaining when the alarm fires, computed over the networks for which a valid alarm is raised. A larger value indicates an earlier warning. We mark a cell by N/A when the sequential test raises no valid alarm in any of the 40 networks, in which case the lead time is undefined.}
    \label{fig:supf}
\end{figure}

Figure~\ref{fig:supf}(a) shows, for each simulation condition and EWS, the fraction of the 40 networks for which the sup-$F$ test succeeds. The figure indicates that the original $V$ and CV are unreliable, in that their performance depends heavily on the simulation condition. In particular, the original variance does not detect steepening with high probability under descending $D$, with success fraction $\le 0.20$ for all four dynamical systems. By contrast, the baseline-referenced variance $V_\Delta$ shows a significant positive steepening in a large majority of the networks for almost all simulation conditions. The success fraction is at least $0.85$ for eight of the ten simulation conditions, and the smallest success fraction across the simulation conditions is $0.60$. The ratio variants $V_{\rm rel}$ and $\mathrm{CV}_{\rm rel}$ behave similarly but somewhat less consistently.

The test whose results are shown in Fig.~\ref{fig:supf}(a) uses all data up to just before the tipping point. This setting is appropriate for retrospectively assessing whether a usable precursor exists. However, a real monitoring system would observe the EWS sequentially as the environmental condition gradually changes toward a potential tipping event. We therefore applied a sequential version of the sup-$F$ test, in which we enlarged the observation window one control-parameter value at a time, starting far from the tipping point and proceeding toward the tipping point. We raise an alarm at the first window in which we detect a significant positive steepening.

We show the fraction of successful alarms for the sequential sup-$F$ test in Fig.~\ref{fig:supf}(b). The results are similar to those for the non-sequential test shown in Fig.~\ref{fig:supf}(a), and are even quantitatively similar for most pairs of EWS and simulation condition. Figure~\ref{fig:supf}(c) shows the lead time at the alarm, defined as the fraction of the home range still remaining when the alarm fires. We averaged the lead time only over the networks for which the sequential test raises a valid alarm before the tipping point, i.e., only over the networks counted in Fig.~\ref{fig:supf}(b). Figure~\ref{fig:supf}(c) indicates that the baseline-referenced EWS variants typically alarm with a large lead time, which is not the case for the original spatial variance $V$. As in the non-sequential and sequential sup-$F$ tests, the ascending double-well dynamics with $u$ as the control parameter and the ascending SIS dynamics remain the hardest cases for the baseline-referenced EWSs.

The results shown in Fig.~\ref{fig:supf}(b) and (c) are further encouraging because the acceleration of $V_\Delta$, $V_{\rm rel}$, and $\mathrm{CV}_{\rm rel}$ is detectable early and online, not only in hindsight. The three panels of Fig.~\ref{fig:supf} suggest that $V_\Delta$ performs better overall than $V_{\rm rel}$ and $\mathrm{CV}_{\rm rel}$. Therefore, we recommend $V_\Delta$ as our final choice, while continuing to examine all three variants in the following analyses as robustness checks.

\subsection{Comparison with a temporal EWS}\label{sub:temporal}

A spatial EWS uses a single snapshot of all nodes, whereas a classical, and more famous, temporal EWS monitors one node over time and therefore requires many temporal samples. To compare the two approaches, we ask how many temporal samples from a single node are needed for the temporal EWS to match the performance of our spatial EWSs. Because aggregating temporal EWSs across multiple nodes generally improves performance~\cite{MacLaren2023JRoySocInterface, masuda2024anticipating, yu2026covariance} but requires an even larger number of samples, we focus in this section on a temporal EWS computed from a single node.

For each simulation condition, network, and control-parameter value, we collected $L=200$ approximately independent samples after equilibration from ten uniformly randomly chosen nodes. For each number of temporal samples $L'\in\{2,3,\dots,200\}$, we computed the single-node temporal variance as the EWS \cite{carpenter2006rising, harris2020early, Scheffer2009Nature} at each control-parameter value of the home range for each of the ten nodes. We then evaluated the same three performance measures as for the spatial EWSs, namely $\tau'$, non-sequential sup-$F$ success, and sequential sup-$F$ success, each averaged over the ten nodes and the 40 networks. We note that our noise strength for the coupled double-well dynamical system, $\sigma=0.1$, is larger than the $\sigma=0.05$ used in Ref.~\cite{masuda2024anticipating}, so we assess both the temporal and spatial EWSs in a noisier, more challenging regime.

\begin{figure}[htbp]
    \centering
    \includegraphics[width=\textwidth]{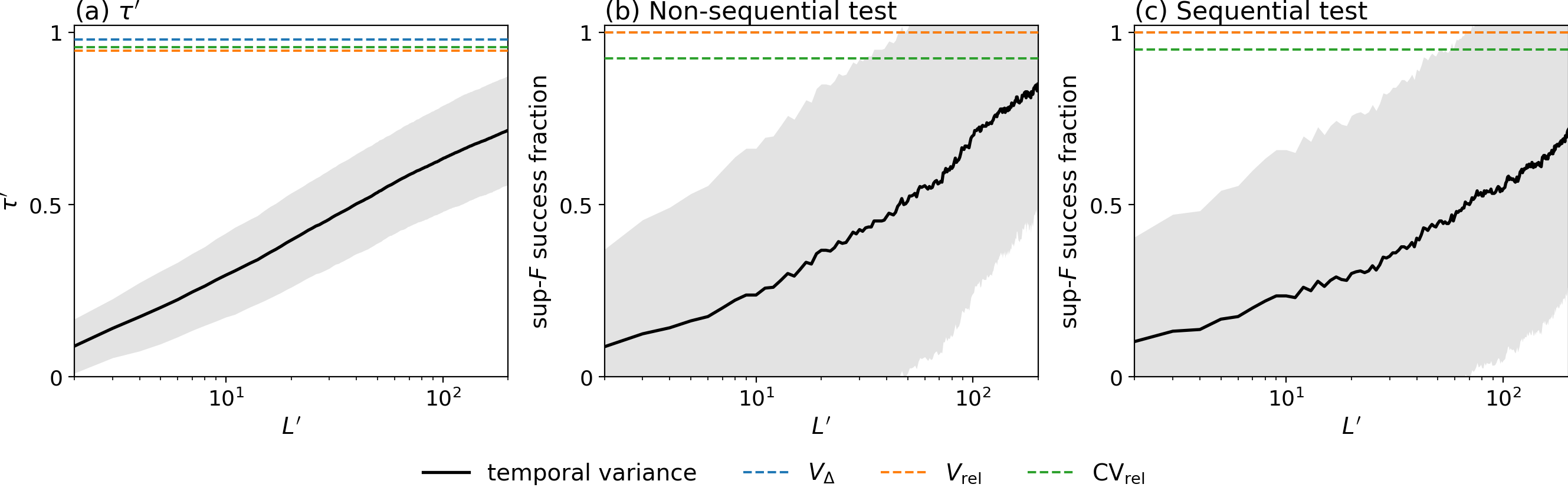}
\caption{Comparison between the single-node temporal variance and the baseline-referenced spatial EWSs. We use the coupled double-well dynamics with descending $D$ as a representative example. (a) Sign-adjusted Kendall's $\tau'$. (b) Non-sequential sup-$F$ success fraction. (c) Sequential sup-$F$ success fraction. The black solid curve shows the average for the temporal variance over the 40 networks and the ten randomly chosen nodes per network, i.e., over $400$ network--node pairs. The shaded area shows one standard deviation above and below the average, also computed over the $400$ network--node pairs. The dashed horizontal lines indicate $V_\Delta$ (blue), $V_{\rm rel}$ (orange), and $\mathrm{CV}_{\rm rel}$ (green), which do not depend on the number of samples, $L'$, used to compute the temporal variance. All three dashed lines overlap at $1$ in (b); the $V_\Delta$ and $V_{\rm rel}$ lines overlap at $1$ in (c).}
    \label{fig:temporal}
\end{figure}

In Fig.~\ref{fig:temporal}, we compare the performance of the temporal variance with varying numbers of samples, $L'$, against our spatial EWSs $V_\Delta$, $V_{\rm rel}$, and $\mathrm{CV}_{\rm rel}$. The figure shows the case of descending $D$ in the coupled double-well dynamics as a representative example. We show the average and standard deviation of the performance measures for the temporal variance by the solid lines and shaded areas, respectively. As expected, for the single-node temporal variance, $\tau'$ (Fig.~\ref{fig:temporal}(a)), the non-sequential sup-$F$ success fraction (Fig.~\ref{fig:temporal}(b)), and its sequential counterpart (Fig.~\ref{fig:temporal}(c)) increase with $L'$, because more samples yield a cleaner temporal estimate. However, even at $L'=200$, the performance measures for the single-node temporal variance remain below those for $V_\Delta$, $V_{\rm rel}$, and $\mathrm{CV}_{\rm rel}$. This behavior is representative: $\tau'$ for the single-node temporal variance and, even more clearly, its sup-$F$ success fraction do not reach those of the spatial $V_\Delta$ within $L'\le200$ in any of the ten simulation conditions. They reach those of $V_{\rm rel}$ and $\mathrm{CV}_{\rm rel}$ only for a small number of simulation conditions (see section~S3 for the results). In other words, within the range of our numerical experiments, a single spatial snapshot of the network carries more early-warning information than a long, temporally resolved record from a single node.

\subsection{Sentinel-node selection}\label{sub:sentinel}

So far, every spatial EWS has used all $N$ nodes. We now ask whether computing the EWS from a subset of nodes, called sentinel nodes, can retain or improve its performance, as has been reported in previous temporal EWS studies \cite{Dakos2014PNAS, Dakos2018EcolInd, MacLaren2023JRoySocInterface, Aparicio2021PNAS, masuda2024anticipating}.
%
%
We keep the same five EWSs, i.e., the original $V$ and CV and the three baseline-referenced variants, $V_\Delta$, $V_{\rm rel}$, and $\mathrm{CV}_{\rm rel}$, and vary the sentinel fraction from $5\%$ to $100\%$; the $100\%$ case is the all-node analysis presented in the previous sections. We use the same three performance measures as before, namely $\tau'$, the non-sequential sup-$F$ detection fraction, and the sequential sup-$F$ detection fraction, each averaged over the 40 networks and the ten simulation conditions. We compare eight rules for choosing the sentinel set, all of which use only the early part of the home range, i.e., the first $\ell=5$ control-parameter values, so that no look-ahead is involved: (a) uniformly at random; (b) nodes with the largest baseline $b_i$; (c) nodes with the smallest $b_i$; (d) nodes with the largest $b_i$ for ascending simulations and the smallest $b_i$ for descending simulations; (e) half of the nodes with the largest $b_i$ and half with the smallest $b_i$; (f) nodes equidistant in the rank of $b_i$; (g) nodes with the smallest early-home-range CV of $x_i$; and (h) nodes with the largest early-home-range CV of $x_i$.

Figure~\ref{fig:sentinel} shows the three performance measures (columns) as functions of the sentinel fraction for each sentinel-node selection rule (rows). We make four observations. First, for every combination of sentinel-node selection rule and performance measure, the three baseline-referenced EWSs (solid lines) lie well above the two original EWSs (dotted lines), confirming that baseline referencing improves the EWSs. Second, $V_\Delta$ is almost always the top performer among the five EWS variants, consistent with the results in the previous sections. Third, for every sentinel-node selection rule and performance measure, the curves are essentially flat or gently increasing with the sentinel fraction, indicating that using more nodes never hurts and that the full network ($100\%$) is at or near the best. In other words, no selection rule yields a clear interior optimum that outperforms the full network. However, a positive finding is that, except for low-performing cases, i.e., low-performing EWSs or low-performing selection rules, spatial EWSs computed from sentinel nodes have almost the same performance as those computed from all nodes, down to a sentinel fraction of approximately $20\%$. This result implies that monitoring only about $20\%$ of the nodes is practically sufficient. Fourth, the selection rule has little effect; the different selection rules perform approximately as well as random sentinel selection (Fig.~\ref{fig:sentinel}(a1)--(a3)).

\begin{figure}[htbp]
    \centering
    \includegraphics[width=\textwidth]{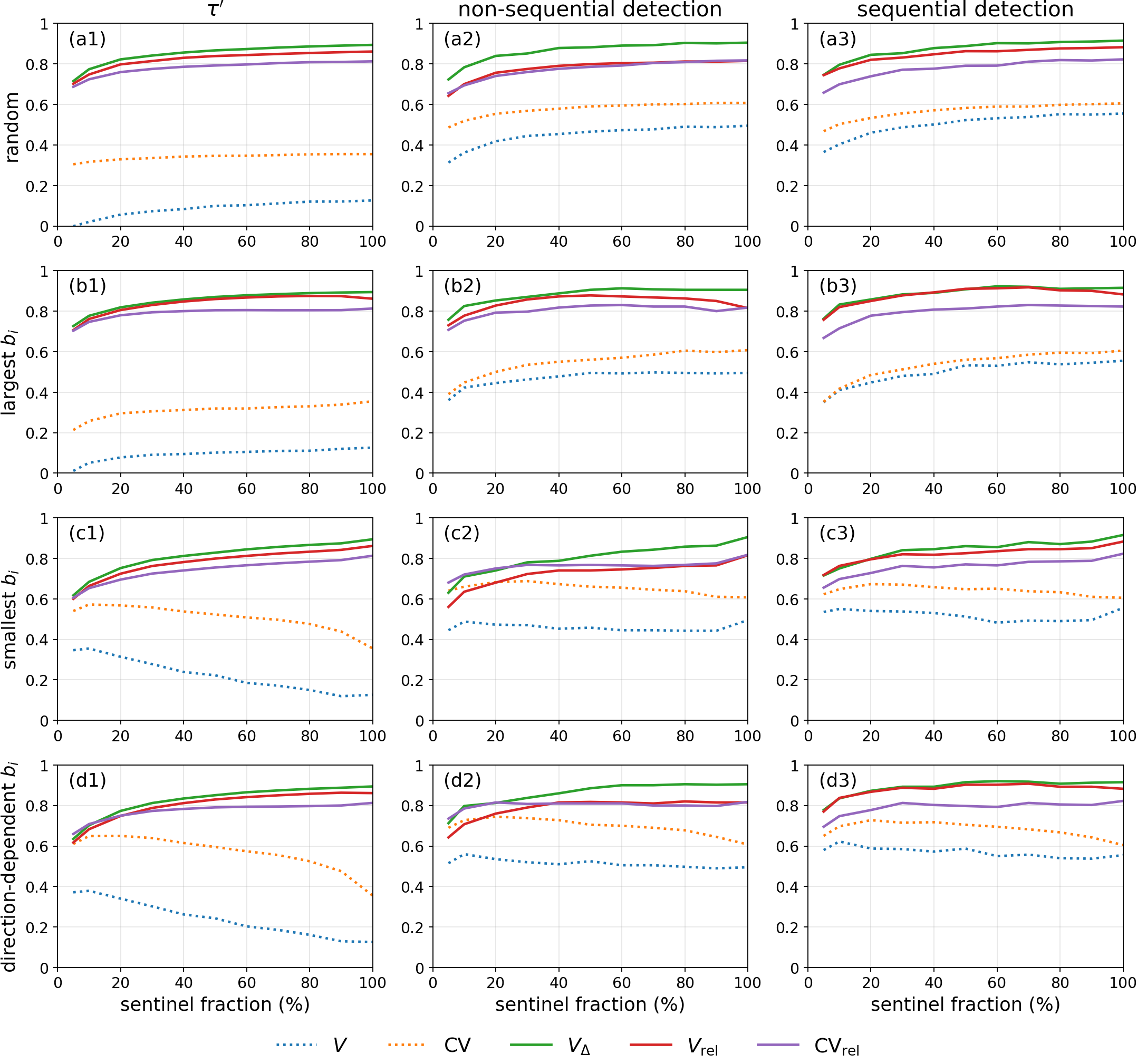}
\caption{Effects of sentinel-node selection on the performance of the spatial EWSs. In each row, the three panels show, from left to right, the three performance measures, namely $\tau'$, the non-sequential sup-$F$ detection fraction, and the sequential sup-$F$ detection fraction, each averaged over the 40 networks and the ten simulation conditions, as a function of the sentinel fraction. Each row corresponds to one sentinel-node selection rule. The five lines in each panel correspond to the five EWSs: the original $V$ and CV and the three baseline-referenced variants, $V_\Delta$, $V_{\rm rel}$, and $\mathrm{CV}_{\rm rel}$. (a1)--(a3) Random sentinel nodes. (b1)--(b3) Largest $b_i$. (c1)--(c3) Smallest $b_i$. (d1)--(d3) Largest $b_i$ for ascending simulations and smallest $b_i$ for descending simulations. (e1)--(e3) Half of the nodes with the largest $b_i$ and half with the smallest $b_i$. (f1)--(f3) Nodes equidistant in the rank of $b_i$. (g1)--(g3) Smallest early-home-range CV of $x_i$. (h1)--(h3) Largest early-home-range CV of $x_i$.}
    \label{fig:sentinel}
\end{figure}
\begin{figure}[htbp]\ContinuedFloat
    \centering
    \includegraphics[width=\textwidth]{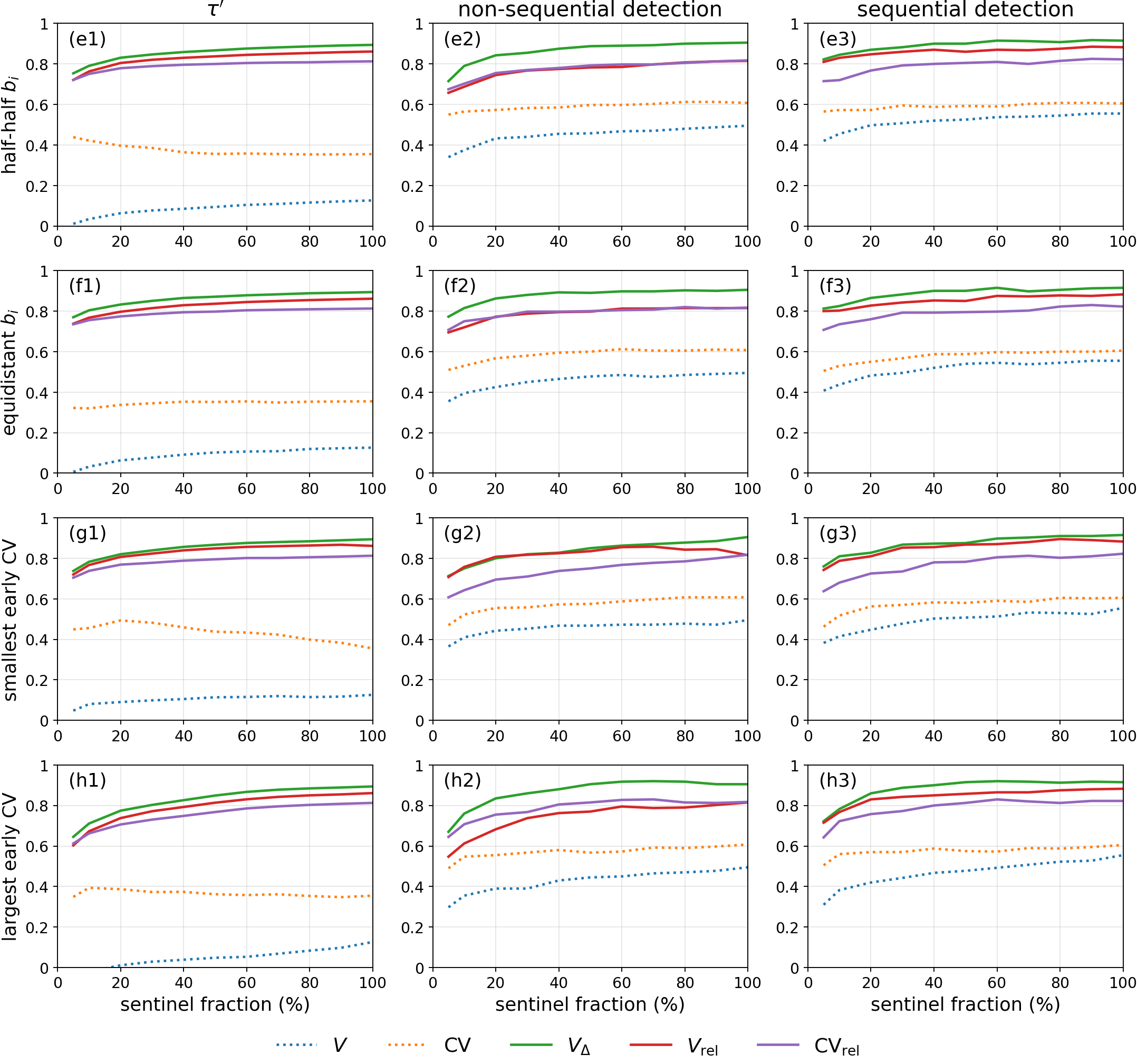}
    \caption{(Continued.) }
\end{figure}

\section{Discussion}

We have developed a baseline-referenced framework for making spatial EWSs more reliable on heterogeneous networks. The central difficulty is that a spatial statistic computed directly from raw node states mixes two effects: static node-to-node heterogeneity induced by network structure and dynamical changes associated with an approaching tipping point. By referencing each node to its own baseline far from the transition, we reduce the former contribution and make the latter more detectable. Across four stochastic dynamical-system models, 40 networks, and ten tipping scenarios, this simple transformation markedly improved variance-based spatial EWSs. In particular, the additive baseline-referenced variance, $V_\Delta$, was the most robust signal. It increased consistently toward tipping points, showed significant positive steepening in the sup-$F$ test, and outperformed a single-node temporal variance within the range of temporal sample sizes tested. We further showed that monitoring roughly 20\% of the nodes often retained nearly the same performance as using all nodes. Together, these results suggest that baseline referencing and sentinel-node sampling can turn spatial EWSs on heterogeneous networks into more robust and practically economical monitoring tools.

Baseline referencing also addresses how one should compare node states when their units or natural scales differ. This issue is common in potential applications of spatial EWSs. In species-interaction dynamics, for example, $x_i$ may represent the biomass of the $i$th animal or plant species, and it is not always clear that the biomasses of different species should be treated as directly comparable quantities \cite{Berlow2004JAnimEcol, GrossFeudel2006PhysRevE, Novak2016AnnuRevEcolEvolSyst}. Similarly, in applications to anticipating depressive symptoms \cite{dablander2023anticipating}, the components of $\{x_i\}$ may be responses to different questionnaire items or continuously monitored physiological signals, again measured on different scales. The multiplicative baseline-referenced variant, which computes a spatial EWS from $\{x_i/b_i\}$ rather than from $\{x_i\}$, has the advantage of being invariant to changes in the unit of individual node states. However, this scale invariance comes at a cost: if some $x_i$ values are close to zero in the early home range, then the corresponding $b_i$ values are also close to zero, and the ratios $x_i/b_i$ can become unstable or artificially large. The additive variant, computed from $\{x_i-b_i\}$, does not particularly suffer from such near-zero baselines, although it is not invariant under node-wise changes of units. Our finding that $V_\Delta$ outperforms $V_{\rm rel}$ and the other variants suggests that, for our numerical data, robustness to near-zero baselines is more important than invariance to unit changes. This interpretation is consistent with the fact that three of the four dynamical systems we used, i.e., the mutualistic species-interaction, SIS, and gene-regulatory models, have $x_i^*=0$ for all $i$ as a stable equilibrium in the absence of dynamical noise, such as the disease-free equilibrium of the SIS model. In empirical applications involving observables with genuinely incomparable units, however, the multiplicative variant may be preferable. We therefore recommend examining both additive and multiplicative baseline-referenced EWSs when applying the method to real-world data.

The weak dependence of the performance of the baseline-referenced variance on the sentinel fraction (shown in Fig.~\ref{fig:sentinel}) can be understood from elementary sampling theory. At a fixed value of the control parameter, baseline referencing removes much of the static node-to-node heterogeneity. Therefore, as a first approximation, the baseline-referenced node states, denoted by $y_i$, can be regarded as samples from a common, control-parameter-dependent distribution whose variance increases as the tipping point is approached. Here, $y_i=x_i^*-b_i$ for $V_\Delta$ and $y_i=x_i^*/b_i$ for $V_{\rm rel}$. If $n$ sentinel nodes are used, the spatial variance is the sample variance
$\hat{V}_n=(n-1)^{-1}\sum_{i=1}^n (y_i-\bar y)^2$.
For independent samples with variance $\sigma^2$ and fourth central moment $\mu_4$, the variance of $V_n$
is equal to $\frac{1}{n}\left(\mu_4-\frac{n-3}{n-1}\sigma^4\right)$
\cite{speed2002tukey, cho2008variance, vegas2012direct}, which reduces to $2\sigma^4/(n-1)$ for a normal distribution. Therefore, the standard deviation of $\hat{V}_n$ scales as $O(n^{-1/2})$. Using only $20\%$ of the nodes then increases the sampling error by a factor of approximately $\sqrt{5}$ relative to using all nodes. However, this increase in the sampling error does not drastically degrade the estimated variance or its increasing trend with the control parameter in practice, whereas the conclusion depends on $N$ (i.e., the number of nodes in the network). This robustness may be because our performance measures duly depend primarily on the monotonicity and acceleration of the EWS across the home range, rather than on the precise pointwise value of the variance at each control-parameter value. Correlations among node states \cite{Clarke2026JPhysComplexity} reduce the effective number of independent samples, but the same qualitative argument applies after replacing $n$ by an effective sample size.

We assessed the control-parameter dependence of our spatial EWSs with the sup-$F$ test rather than with a bifurcation-specific model-comparison method (e.g., \cite{hessler2022bayesian}) because the latter requires specifying a functional form for the EWS near the transition. Modern approaches that statistically compare a bifurcating model against a non-bifurcating null model \cite{hessler2022bayesian, ditlevsen2023warning, masuda2026tipmoc} typically exploit theoretically predicted behavior of an EWS near a bifurcation, often a power-law divergence. Such laws are usually derived for ensemble statistics of a stochastic dynamical system, such as the variance of a state variable. Temporal sample statistics approximate these ensemble quantities under stationarity and ergodicity assumptions. By contrast, a spatial EWS is a cross-sectional statistic computed across nodes of a heterogeneous network. Even after baseline referencing, it is not generally clear whether such a spatial statistic obeys the same law as the corresponding temporal or ensemble statistic when the tipping point is approached (see \cite{Clarke2026JPhysComplexity} for an analysis for spatial EWSs in $\mathbb{R}$). Establishing such theory for spatial EWSs on networks is an important direction for future work. The sup-$F$ test avoids imposing a bifurcation-specific scaling form; it only asks whether the EWS shows a significant positive steepening as the tipping point is approached. This makes it a comparatively assumption-light diagnostic that can be applied across different dynamical systems and tipping scenarios.

Several limitations point to useful next steps. First, the present study is computational, and we have used only stylized dynamical system models. Empirical validation is left for future work. Second, we focused on static, undirected, and unweighted networks. Extending the present approach to temporal, multilayer, directed, or weighted networks, or to systems with higher-order interactions such as hypergraphs, would broaden its applicability. However, our recommended baseline-referenced variance, $V_\Delta$, does not require knowledge of the network structure, i.e., the adjacency matrix. Third, although random sentinel-node selection was sufficient to retain much of the performance in our simulations, it remains unclear whether more predictive sentinel nodes can be identified systematically. Addressing these questions is considered to further help turn baseline-referenced spatial EWSs from a robust numerical prescription into a deployable framework for anticipating tipping points in real multivariate data from interconnected systems.

\section{Methods}\label{methods}

\subsection{Dynamics}\label{dynamics}

We consider four stochastic dynamical-system models on networks: coupled double-well dynamics, susceptible--infectious--susceptible (SIS) dynamics, mutualistic species-interaction dynamics, and gene-regulatory dynamics. The coupled double-well dynamics is given by Eq.~\eqref{eq:doublewell}. When we vary $u$, we set $D = 0.05$. When we vary $D$, we set $u = 0$ in ascending simulations and $u = -5$ in descending simulations; the negative stress in the descending case is necessary to destabilize the upper equilibrium so that a downward transition occurs as $D$ decreases. We initialize $x_i = 1$ for all $i$ in ascending simulations, so that each node starts in its lower state unless the dynamical noise is excessively large. Similarly, we initialize $x_i = 5$ for all $i$ in descending simulations, so that each node starts in its upper state. We classify node $i$ as being in its lower state if $x_i < r_2$ and in its upper state otherwise.

The mutualistic species-interaction dynamics is given by \cite{gao2016universal}
\begin{equation}\label{mutualistic}
dx_i = \left[ B + x_i\left(1 - \frac{x_i}{K}\right) \left(\frac{x_i}{C} - 1\right) + D \sum_{j=1}^N A_{ij} \frac{x_i x_j}{\tilde{D} + E x_i + H x_j} + u \right] dt + \sigma d\xi_i,
\end{equation}
where $x_i$ represents the abundance of the $i$th species, $B$ is a constant immigration rate, $K$ is the carrying capacity, $C$ is the Allee threshold, and $\tilde{D}$, $E$, and $H$ modulate the interaction term $x_i x_j$. The term $x_i\left(1 - x_i/K\right) \left(x_i/C - 1\right)$ represents logistic growth with an Allee effect. Following \cite{gao2016universal}, we set $B = 0.1$, $K = 5$, $C = 1$, $\tilde{D} = 5$, $E = 0.9$, $H = 0.1$, and $\sigma = 0.001$. Furthermore, we set $D = 0.05$ if the control parameter is $u$, and $u = -5$ if the control parameter is $D$. Similar to the coupled double-well dynamics, the mutualistic interaction model has two stable states, namely, a lower state with $x_i = 0$ and an upper state with $x_i = K$, in the absence of coupling and noise. We initialize the model in the upper state by setting $x_i = 6$ for all nodes $i$, representing an established population. Throughout the simulation, we classify node $i$ as being in the upper state if $x_i > C$ and in the lower state otherwise. To prevent nonphysical values, we set any negative value of $x_i$ to zero as soon as it appears in this and the following two models.

A stochastic ODE variant of the SIS model on networks is given by
\begin{equation}\label{sis}
dx_i = \left[ -\mu x_i + D \sum_{j=1}^N A_{ij} (1 - x_i)x_j \right] dt + \sigma d\xi_i,
\end{equation}
where $x_i$ represents the probability that the $i$th node is infectious and $1-x_i$ is the probability that it is susceptible. Parameters $D$ and $\mu$ are the infection and recovery rates, respectively. We set $\mu = 1$ and $\sigma = 0.001$. In the absence of noise, the system admits a disease-free equilibrium at $x_i = 0$ for all $i$, which remains stable when the infection rate $D$ is below an epidemic threshold; when $D$ is larger than this threshold, the system has a stable endemic equilibrium with $x_i > 0$ for all $i$, provided that the network is connected \cite{pastor2015epidemic}. However, when noise is present, small but nonzero values of $x_i$ can occur for any $D \geq 0$ because of fluctuations. In the ascending simulations, we initialize the simulations near the disease-free state, with $x_i = 0.001$ for all $i$. In the descending simulations, we initialize them in a high-prevalence state, with $x_i = 0.999$ for all $i$. Following \cite{maclaren2025applicability}, we classify node $i$ as being in its lower state if and only if $x_i < 5\sigma$ in this model and in the gene-regulatory model described next.

A model of gene-regulatory dynamics on networks is given by \cite{gao2016universal}
\begin{equation}
\label{genereg}
dx_i = \left( -B x_i^f + D \sum_{j=1}^N A_{ij} \frac{x_j^h}{1 + x_j^h} + u \right) dt + \sigma d\xi_i,
\end{equation}
where $x_i$ represents the expression level of the $i$th gene. The term $-B x_i^f$ represents self-degradation or negative autoregulation of the $i$th gene's expression. Following \cite{gao2016universal}, we set $B = 1$, $f = 1$, $h = 2$, and $\sigma = 0.001$. Furthermore, we fix $D = 1$ if the control parameter is $u$, and $u = 0$ if the control parameter is $D$. In the absence of noise, the model has an equilibrium at $x_i = 0$ for all $i$. This inactive state is stable when $u$ or $D$ is sufficiently small. An active state with $x_i > 0$ for all $i$ exists when $u$ or $D$ is sufficiently large. We initialize the simulations near the upper state, with $x_i = 2$ for all $i$, to study the loss of resilience of the active state.

For the coupled double-well, mutualistic species-interaction, and gene-regulatory models, we use $D\in\mathbb{R}_{\geq 0}$ and $u \in \mathbb{R}$ as control parameters, i.e., bifurcation parameters. For the SIS model, we use $D$ as the sole control parameter because a uniform stress parameter, $u$, is not physically relevant for epidemic dynamics.

\subsection{Networks}\label{networks}

We used 40 networks, comprising 34 empirical networks and six synthetic networks. Most of these networks, i.e., 36 of them, are the same as those used in our previous study \cite{maclaren2025applicability}. To assess scalability and robustness on larger networks, we added four larger empirical networks. For each network, we used the largest connected component and treated it as an undirected and unweighted network, removing any multiple edges. We provide details of each network in Section~S1.

\subsection{Simulations}\label{simulations}

Each simulation setting, which we refer to as a simulation condition, is a combination of the dynamical-system model (coupled double-well, mutualistic species-interaction, SIS, or gene-regulatory dynamics), control parameter ($D$ or $u$), and direction in which the control parameter is varied (ascending or descending). For example, in Fig.~\ref{fig:existing}(a)--(e), the simulation condition is coupled double-well dynamics with $D$ as the descending control parameter. In what follows, we describe the simulation protocol for the ascending case in detail.

In each simulation, we use the Euler--Maruyama scheme with a fixed time step. Most simulations use $\Delta t = 0.01$, whereas some use $\Delta t = 0.005$. For the four largest networks, we use a smaller time step, down to $\Delta t = 0.001$, so that the equilibrated state is numerically converged. We report the integration time, time step, and control-parameter range for each simulation in a supplementary file explained in section S4. We regard the state at the final time, denoted by $x_i^*$ for each node $i$, as an equilibrated sample in the presence of dynamical noise. One sample of $x_i^*$ is recorded from each node for computing the spatial EWSs at the current value of the control parameter. We then increment or decrement the control parameter, depending on whether the simulation is ascending or descending, respectively, by a small amount and repeat the simulation starting from the same initial condition. We repeat this procedure across $100$ uniformly spaced values of the control parameter. These $100$ control-parameter values define the simulation range \cite{maclaren2025applicability}.

Within the simulation range, a tipping event, i.e., a bifurcation of the dynamical system, may occur. Because our goal is to anticipate tipping points, equilibrated states after a tipping event are not informative for early warning. Thus, we compute EWSs only for control-parameter values at which all $x_i^*$ remain close to their initial state, i.e., before the first tipping event. We refer to this restricted region of the control parameter as the home range \cite{maclaren2025applicability}. The home range is a subset of the simulation range.

We have confirmed that the simulation range contains at least one transition, so that the home range is well defined. For most combinations of dynamics and networks, we adopt the simulation ranges previously identified in \cite{maclaren2025applicability}. For additional networks not considered in that study and in some exceptional cases, we determined the simulation ranges by trial and error (section S4). We report all simulation ranges and home ranges used in the present study in the aforementioned supplementary file.

\subsection{Early warning signals}\label{earlywarningsignals}

We evaluate five classical spatial EWSs, each of which requires only a single sample of $x_i^*$ per node at a given value of the control parameter. We define them here for the full set of all $N$ nodes. We denote the sample mean by $\overline{x} = N^{-1}\sum_{i=1}^{N} x_i$.

The spatial variance is given by \cite{guttal2009spatial, eby2017alternative}
\begin{equation}
    V = \frac{1}{N-1}\sum_{i = 1}^{N} (x_i - \overline{x})^2.
\end{equation}

The coefficient of variation (CV) is defined by \cite{dai2013slower, litzow2008increased, rindi2018experimental}
\begin{equation}
    \text{CV} = \frac{\sqrt{V}}{\overline{x}}.
\end{equation}

The $k$th central moment of $\{ x_1, \ldots, x_N \}$ is $m_k = N^{-1} \sum_{i = 1}^{N} (x_i - \overline{x})^k$. The sample skewness, which is widely used as a spatial EWS \cite{guttal2009spatial, buelo2018modeling}, is the scaled third central moment, given by
\begin{equation}
    g_1 = \frac{m_3}{m_{2}^{3/2}}.
\end{equation}
Sample skewness measures the asymmetry of a distribution, indicating whether extreme values are more likely to occur on the right (positive skew) or left (negative skew) side of the mean. As a dynamical system approaches a tipping point, skewness may increase, becoming more positive, or decrease, becoming more negative, depending on the simulation direction. To remove this dependence on the simulation direction, we use a sign-adjusted skewness, $g'_1$, where $g'_1 \equiv g_1$ for ascending simulations and $g'_1 \equiv -g_1$ for descending simulations \cite{maclaren2025applicability}. With this convention, an increase in $g'_1$ is consistently associated with proximity to a tipping point.

The sample kurtosis is the scaled fourth central moment \cite{litzow2017indications, buelo2018modeling}, given by
\begin{equation}
    g_2 = \frac{m_4}{m_{2}^{2}}.
\end{equation}
It measures the propensity of a distribution to produce extreme values. For a normal distribution, $g_2$ converges to $3$ as the sample size tends to infinity.

Moran's $I$ is given by \cite{legendre1989spatial, Dakos2010TheorEcol, okabe2015spatial}
\begin{equation}\label{eq:moran}
I_{M}=\frac{N}{W}\cdot
\frac{\sum_{i=1}^{N}\sum_{j=1}^{N} A_{ij}\,(x_i-\overline{x})(x_j-\overline{x})}
{\sum_{i=1}^{N}(x_i-\overline{x})^2},
\end{equation}
where $W=\sum_{i=1}^{N}\sum_{j=1}^{N} A_{ij}$. Moran's $I$ measures spatial autocorrelation along network edges; large positive values indicate that adjacent nodes tend to have similar states. Unlike the other four EWSs, Moran's $I$ uses the network structure, i.e., the adjacency matrix.

For the spatial variance and CV, we also use sentinel-node versions, in which we compute the statistic over a subset of $n$ ($\le N$) sentinel nodes instead of all $N$ nodes. This amounts to restricting the sums above to the sentinel set and replacing $N$ by $n$.

\subsection{Kendall's $\tau$}\label{kendallstau}

To evaluate the effectiveness of EWSs, we use Kendall's rank correlation coefficient, denoted by $\tau$, which has been widely used in EWS studies \cite{kefi2014early, Dakos2010TheorEcol, chen2022practical}. We compute Kendall's $\tau$ between the control-parameter values in the home range and the corresponding EWS values.

In practice, we use the sign-adjusted variant, $\tau'$, to remove the dependence of the sign of $\tau$ on the simulation direction \cite{maclaren2025applicability}. We set $\tau' = \tau$ in ascending simulations and $\tau' = -\tau$ in descending simulations. This adjustment ensures that larger $\tau'$ values correspond to better EWS performance regardless of the simulation direction. Note that $\tau, \tau' \in [-1, 1]$.

\subsection{sup-$F$ test}\label{sub:supfmethod}

To test whether an EWS accelerates toward the tipping point, we use the sup-$F$ test for a structural change at an unknown breakpoint \cite{quandt1960tests, andrews1993tests, zeileis2002strucchange}. For a given EWS evaluated over the home range, we orient the control parameter so that it increases toward the tipping point. For each candidate breakpoint, we fit two separate straight lines, one to the data before the breakpoint and one to the data after it, allowing both the intercept and the slope to differ between the two segments. The fitted curve is not constrained to be continuous at the breakpoint. We then compute the $F$ statistic for the null hypothesis that the two segments share a common intercept and a common slope, which is the classical Chow statistic for a break at a prescribed point \cite{chow1960tests}. Because the breakpoint is unknown, we take the supremum of this $F$ statistic over all candidate breakpoints \cite{quandt1960tests, andrews1993tests, zeileis2002strucchange}. We consider only breakpoints that leave at least $\max\left(4, \lceil 0.15 q \rceil\right)$ of the $q$ points in each segment, where $15\%$ trimming is a standard choice \cite{andrews1993tests}. A small $p$-value indicates a significant change in the regression coefficients, i.e., in the intercept, the slope, or both, at the maximizing breakpoint.

We obtain $p$-values by Monte Carlo simulation under the null hypothesis of a single linear trend with independent Gaussian noise. Because the sup-$F$ statistic is invariant to the intercept and the slope of the underlying linear trend and to the scale of the noise, its null distribution depends only on the number of control-parameter values, $q$. Therefore, for each $q$, we generate $\overline{B}=2000$ surrogate series, each consisting of $q$ independently drawn standard normal variables, i.e., a zero trend and unit variance, placed at the same equally spaced control-parameter values. We then compute the sup-$F$ statistic for each surrogate series. The value of the noise standard deviation is immaterial because of the scale invariance. The $p$-value of an observed sup-$F$ statistic, denoted by $F$, is its right-tailed rank among the surrogate statistics, $p=\big(1+\#\{b': F^\ast_{b'}\ge F\}\big)/(\overline{B}+1)$,
where $F^\ast_{b'}$ is the sup-$F$ statistic of the $b'$th surrogate. The independence assumption underlying the surrogates is appropriate here because the equilibrium at each control-parameter value is computed in an independent simulation, so the EWS values are independent across control-parameter values under the null model. We classify the result as a positive steepening when $p<0.05$ and, at the breakpoint maximizing the $F$ statistic, the fitted slope is non-negative before the breakpoint and larger afterward.

The procedure just described is non-sequential. In other words, it uses the entire home range at once and asks, in hindsight, whether a positive steepening occurred. We also run a more stringent, sequential version of the sup-$F$ test that mimics online monitoring. Starting from the control-parameter value farthest from the tipping point, we enlarge the observation window one control-parameter value at a time. At each window, we compute the sup-$F$ statistic over the data observed so far and raise an alarm at the first window whose statistic exceeds a threshold and whose detected slope change is a positive steepening. Repeatedly testing on growing windows creates a multiple-comparison problem. Because the successive windows overlap heavily, the corresponding tests are strongly dependent. A Bonferroni correction remains valid under such dependence but would be unnecessarily conservative. Instead, we calibrate the complete monitoring procedure by Monte Carlo simulation under the same null model. For each surrogate series, we run the identical growing-window procedure and record the maximum sup-$F$ statistic over the monitoring path. We set the alarm threshold to the empirical $(1-p')$ quantile of these pathwise maxima, with $p'=0.05$. This calibration targets a family-wise false-alarm probability of $p'$ under the surrogate null model.

\subsection{Single-node temporal early warning signal}\label{sub:temporalmethod}

To compare our spatial EWSs against a classical temporal EWS, we computed the temporal variance of a single node's state as a function of the control parameter. For each simulation condition and network, and at each control-parameter value in the home range, we continued simulating the stochastic dynamics after equilibration. We recorded $L=200$ samples of the state $x_i(t)$ of each of ten uniformly randomly chosen nodes $i$, with consecutive samples separated by one time unit for the coupled double-well, SIS, and gene-regulatory dynamics and by $0.1$ time units for the mutualistic species-interaction dynamics~\cite{masuda2024anticipating}.

For each number of samples $L'\in\{2,3,\dots,200\}$, we computed, for each node $i$, the single-node temporal variance at a given control-parameter value as the sample variance of the first $L'$ samples of $x_i(t)$. This procedure yields a series of temporal-variance values across the control-parameter values in the home range. We applied the same three performance measures used for the spatial EWSs, i.e., $\tau'$, the non-sequential sup-$F$ test, and the sequential sup-$F$ test, to this series. We report, as a function of $L'$, the $\tau'$ averaged over the ten nodes and the 40 networks, and the sup-$F$ success fractions, i.e., the fraction of the $10\times40$ node--network pairs for which the sup-$F$-based procedure identifies a positive steepening.

\subsection{Sentinel-node selection}\label{sub:sentinelmethod}

When a spatial EWS is computed from a subset of $n$ sentinel nodes, we set $n=\max(3,\lceil \overline{f} N\rceil)$ for a target fraction $\overline{f}$ of the $N$ nodes. 
%
%
Every selection rule uses only the baseline $b_i$ (Eq.~\eqref{eq:baseline}) or the temporal CV, denoted by $\gamma_i$, i.e., the standard deviation divided by the mean, computed from the first $\ell=5$ home-range values of $x_i^*$ for node $i$. We thus fix the sentinel-node set after the first $\ell$ control-parameter values and monitor the EWSs computed from these sentinel nodes as the dynamical system approaches the tipping point.

We use the following eight rules. (a) Random: $n$ nodes drawn uniformly at random, averaging over five independent draws per network. (b) Largest and (c) smallest $b_i$: the $n$ nodes with the largest or smallest $b_i$, respectively. (d) Direction-dependent: the largest $b_i$ for ascending simulations and the smallest $b_i$ for descending simulations. (e) Half--half: the $\lceil n/2\rceil$ nodes with the largest $b_i$ together with the $\lfloor n/2\rfloor$ nodes with the smallest $b_i$. (f) Equidistant: we sort the nodes by $b_i$ and take $n$ of them at equally spaced ranks, including the nodes with the largest and smallest $b_i$. (g) Smallest and (h) largest CV: the $n$ nodes with the smallest or largest early-home-range CV, i.e., $\gamma_i$, respectively. For each sentinel set, we evaluate the five EWSs and the three performance measures.

\section*{Acknowledgments}

We thank Esteban Vargas Bernal for discussion about the sup-$F$ test.
N.M. acknowledges financial support by the Japan Science and Technology Agency (JST) Moonshot R\&D (under Grant No. JPMJMS2021), the National Science Foundation (under grant no.\,2204936), and JSPS KAKENHI (under grant nos.\,JP 23H03414, 24K14840, and 24K03013).
During the preparation of this manuscript, the authors used Claude Opus 4.7 for language editing and code development assistance. The authors manually verified all code and manuscript text.


\clearpage
\setcounter{section}{0}
\setcounter{figure}{0}
\setcounter{table}{0}
\setcounter{equation}{0}
\setcounter{page}{1}
\renewcommand{\thesection}{S\arabic{section}}
\renewcommand{\thefigure}{S\arabic{figure}}
\renewcommand{\thetable}{S\arabic{table}}
\renewcommand{\theequation}{S\arabic{equation}}
\renewcommand{\theHsection}{SI.\arabic{section}}
\renewcommand{\theHfigure}{SI.\arabic{figure}}
\renewcommand{\theHtable}{SI.\arabic{table}}
\renewcommand{\theHequation}{SI.\arabic{equation}}
\renewcommand{\refname}{Supplementary References}

\begingroup
\setcounter{footnote}{0}%
\null
\vskip 2em%
\begin{center}%
  {\LARGE \textbf{Supplementary Material for: \\
Baseline-referenced spatial early warning signals for tipping points on heterogeneous networks} \par}%
  \vskip 1.5em%
  {\large
   \lineskip .5em%
   \begin{tabular}[t]{c}%
Tharusha Bandara, Shilong Yu, Naoki Masuda
   \end{tabular}\par}%
  \vskip 1em%
  {\large }%
\end{center}%
\par
\vskip 1.5em%
\endgroup

\section{Networks}

In our numerical simulations, we used 40 networks, comprising 34 empirical networks and six synthetic networks. Following theoretical ecology studies that test spatial early warning signals (EWSs) \cite{Dakos2010TheorEcol-SI,kefi2014early-SI}, one of the six synthetic networks is a square lattice with periodic boundary conditions. The other five synthetic networks, i.e., GKK, ER, WS, BA, and HK, are the same as those used in our previous study \cite{maclaren2025applicability-SI}. Among the 34 empirical networks, the 30 smaller networks are also the same as those used in our previous study \cite{maclaren2025applicability-SI}. We added four larger empirical networks to assess the generality of our main results to larger networks. For each network, we extracted the largest connected component, removed any multiple edges, and converted the network into an undirected and unweighted network. We provide a brief description and the numbers of nodes and edges for all networks in Table~\ref{si-networks}.

We obtained the empirical networks from the KONECT repository \cite{kunegis2013konect-SI} or the Netzschleuder collection \cite{peixoto2020netzschleuder-SI}. We generated the synthetic networks using the \texttt{igraph} package \cite{csardi2024igraph-SI}, except for the Holme--Kim (HK) model network, which we generated using NetworkX \cite{hagberg2007exploring-SI}.

\begin{longtable}{@{}lrrp{0.68\linewidth}@{}}
\caption{Networks used. $N$: number of nodes, $M$: number of edges.
\label{si-networks}}\\
\toprule
Network & $N$ & $M$ & Notes \\
\midrule
\endfirsthead

\caption[]{Networks used (continued).}\\
\toprule
Network name & $N$ & $M$ & Notes \\
\midrule
\endhead

\midrule
\multicolumn{4}{r}{Continued on next page} \\
\endfoot

\bottomrule
\endlastfoot

Montreal & 29 & 75 & Network describing alliances and rivalries among street gangs operating in Montreal, Quebec \cite{descormiers2011alliances-SI}. The original data did not distinguish between alliance and rivalry ties.\\

Chesapeake & 39 & 170 & Trophic ecosystem network representing major biological components of Chesapeake Bay \cite{baird1989seasonal-SI}. Nodes correspond to ecological groups such as phytoplankton or fish larvae. Edges indicate carbon transfer through feeding relationships. \\

Windsurfer & 43 & 336 & Interpersonal contact network among windsurfers on a beach in southern California, USA \cite{freeman1988human-SI}. \\

Geographic & 49 & 107 & Geographic adjacency network among U.S. states and territories \cite{knuth2005art-SI}. Nodes represent states or territories. Edges connect pairs of states or territories that share a border. \\

Catlins & 59 & 110 & Freshwater food web from the Catlins region \cite{thompson2003impacts-SI}. Nodes represent species or taxonomic groups. Edges represent observed predator--prey interactions. \\

Dolphin & 62 & 159 & Social interaction network of bottlenose dolphins \cite{lusseau2003bottlenose-SI}. Nodes represent individual dolphins. Edges represent frequent associations between pairs of animals. \\

Terrorist & 64 & 243 & Contact network among individuals involved in the 2004 train bombing in Madrid, Spain \cite{hayes2006connecting-SI}.\\

Drug interaction & 75 & 181 & Network constructed from medical records in Blumenau, Brazil \cite{correia2019city-SI}. Nodes represent pharmaceutical drugs. Edges indicate documented interactions between medications. \\

Contact & 75 & 114 & Sexual contact network among individuals in Iceland \cite{haraldsdottir1992preliminary-SI}. Edges correspond to reported sexual partnerships. \\

GKK & 96 & 300 & Synthetic network generated using the Goh--Kahng--Kim model \cite{goh2001universal-SI}. Edge probabilities depend on node fitness values, producing a heterogeneous degree distribution. We set $\alpha = 1$, $N = 100$, and $M = 300$.\\

ER & 100 & 249 & An instance of the Erd\H{o}s--R\'enyi model. Each node pair forms an edge with probability $0.05$. \\

WS & 100 & 400 & Small-world network generated using the Watts--Strogatz model \cite{watts1998collective-SI}. The initial network is a periodic one-dimensional lattice in which each node is adjacent to its four nearest neighbors, followed by random rewiring with probability $0.02$. \\

BA & 100 & 197 & Scale-free network generated using the Barab\'asi--Albert model with parameter $m = 2$ \cite{albert1999emergence-SI}. The construction begins from a fully connected seed graph with three nodes. \\

HK & 100 & 196 & Network generated using the Holme--Kim extension of the Barab\'asi--Albert model with $m = 2$, designed to increase the abundance of triangles \cite{holme2002growing-SI}. The initial network is a chain network with three nodes.\\

Lattice & 100 & 200 & Two-dimensional square lattice consisting of $10 \times 10$ nodes with connections to nearest neighbors under periodic boundary conditions. \\

Canton & 109 & 717 & Freshwater ecological food web representing trophic relationships among species in the Canton ecosystem \cite{thompson2003impacts-SI}. As in the Catlins network, nodes represent taxa, and edges indicate feeding interactions.\\

Gene fusion & 110 & 124 & Network representing gene fusion events identified in human cancers \cite{hoglund2006gene-SI}. Nodes correspond to genes. Edges represent documented fusions between them.\\

Word & 112 & 425 & Network of common adjectives and nouns appearing in the novel \textit{David Copperfield} by Charles Dickens \cite{newman2006finding-SI}. Two words are adjacent if they occur adjacently at least once in the text.\\

Football & 115 & 613 & Network of U.S. college American football games \cite{girvan2002community-SI}. Nodes are collegiate football teams in the United States. Two nodes are adjacent if the corresponding teams played a game during the 2000 season.\\

Physician & 117 & 465 & Social network of physicians in four towns in Illinois, USA \cite{coleman1957diffusion-SI}. An edge indicates that one physician named another as a friend, discussion partner, or source of advice.\\

Student & 141 & 256 & Cooperation network among university students \cite{fire2012predicting-SI}.\\

Protein & 161 & 209 & Biological interaction network \cite{beuming2005pdzbase-SI}. Nodes correspond to proteins. Edges indicate protein--protein interactions.\\

Email & 167 & 3250 & Communication network derived from email exchanges within a manufacturing company \cite{michalski2011matching-SI}. Nodes represent email accounts. Edges indicate that at least one email was exchanged between the corresponding accounts.\\

Village & 187 & 431 & Advice-seeking network from a rural Ugandan community \cite{chami2017social-SI}. Nodes represent households. An edge indicates that a member of one household identified someone from another household as a trusted source of advice.\\

Jazz player & 198 & 2742 & Collaboration network of jazz musicians \cite{gleiser2003community-SI}. Each node represents a musician. Two nodes are adjacent if the corresponding musicians performed together in the same band or ensemble.\\

Flamingo software & 228 & 491 & Software dependency network from the Flamingo project \cite{vsubelj2011community-SI}. Nodes represent classes in an object-oriented codebase. Edges represent dependency relationships between classes.\\

\textit{E. coli} & 328 & 456 & Transcriptional regulatory network of the bacterium \textit{Escherichia coli} \cite{shen2002network-SI}. Nodes are operons, i.e., gene clusters. Edges represent regulatory relationships between operons.\\

Transportation & 369 & 430 & Network of stations in the London Underground, UK \cite{de2014navigability-SI}. Two stations are adjacent if they are directly connected by an underground line.\\

Coauthorship & 379 & 914 & Scientific collaboration network among researchers studying network science \cite{newman2006finding-SI}. Nodes represent authors. Edges indicate coauthorship of at least one publication.\\

Wikipedia user & 404 & 734 & Interaction network among contributors to the Haitian Creole Wikipedia \cite{sun2016predicting-SI}. Two users are adjacent if one posted a message on the other user's talk page.\\

Proximity & 410 & 2765 & Face-to-face contact network at a museum display \cite{isella2011s-SI}. Nodes represent museum visitors. Two nodes are adjacent if the corresponding visitors were sufficiently close to each other in physical space during their visit.\\

\textit{C.\ elegans} metabolic & 453 & 2025 & Metabolic network of the nematode \textit{Caenorhabditis elegans} \cite{jeong2000large-SI}. Nodes represent biochemical entities such as enzymes, metabolites, or transient complexes. Edges indicate participation in the same biochemical reaction.\\

\textit{C.\ elegans} neuronal & 460 & 1432 & Neural connectome of \textit{C.\ elegans} \cite{cook2019whole-SI}. Nodes correspond to neurons. Edges represent synaptic connections between neurons.\\

\textit{S.\ cerevisiae} & 664 & 1065 & Regulatory network of the yeast \textit{Saccharomyces cerevisiae} \cite{milo2002network-SI}. Nodes represent regulatory units. Edges represent regulatory interactions between them.\\

Product & 774 & 1779 & Network of exported products \cite{hidalgo2007product-SI}. Nodes are economic products. Two nodes are adjacent if the corresponding products are sufficiently similar in terms of the countries that export them.\\

JUNG software & 879 & 2047 & Software dependency network derived from the JUNG framework \cite{vsubelj2011community-SI}. As in the Flamingo software network, nodes correspond to classes, and edges indicate dependency relationships between classes.\\

U. Rovira i Virgili & 1133 & 5451 & Email communication network from Universitat Rovira i Virgili in Tarragona, Spain \cite{guimera2003self-SI}. Nodes represent users. An edge indicates that two users exchanged at least one email.\\

US power grid & 4941 & 6594 & Infrastructure network representing the Western States power grid of the USA \cite{watts1998collective-SI}. Nodes correspond to generators, substations, or transformers. Edges represent transmission lines connecting them.\\

Route views & 6474 & 12572 & Snapshot of the Internet at the autonomous-system level, inferred from Border Gateway Protocol routing tables collected by the University of Oregon Route Views project. Each node is an autonomous system. An edge indicates a direct routing relationship between two autonomous systems.\\

Erd\H{o}s collaboration & 6927 & 11850 & Collaboration network centered on the mathematician Paul Erd\H{o}s. Nodes represent researchers who are directly or indirectly connected to Erd\H{o}s through coauthorship. Edges represent coauthorship relationships.\\

\end{longtable}

\newpage

\section{Kendall's $\tau$ for spatial EWSs over the second half of the home range}\label{sec:si-half}

\begin{figure}[htbp]
\centering
\includegraphics[width=\textwidth]{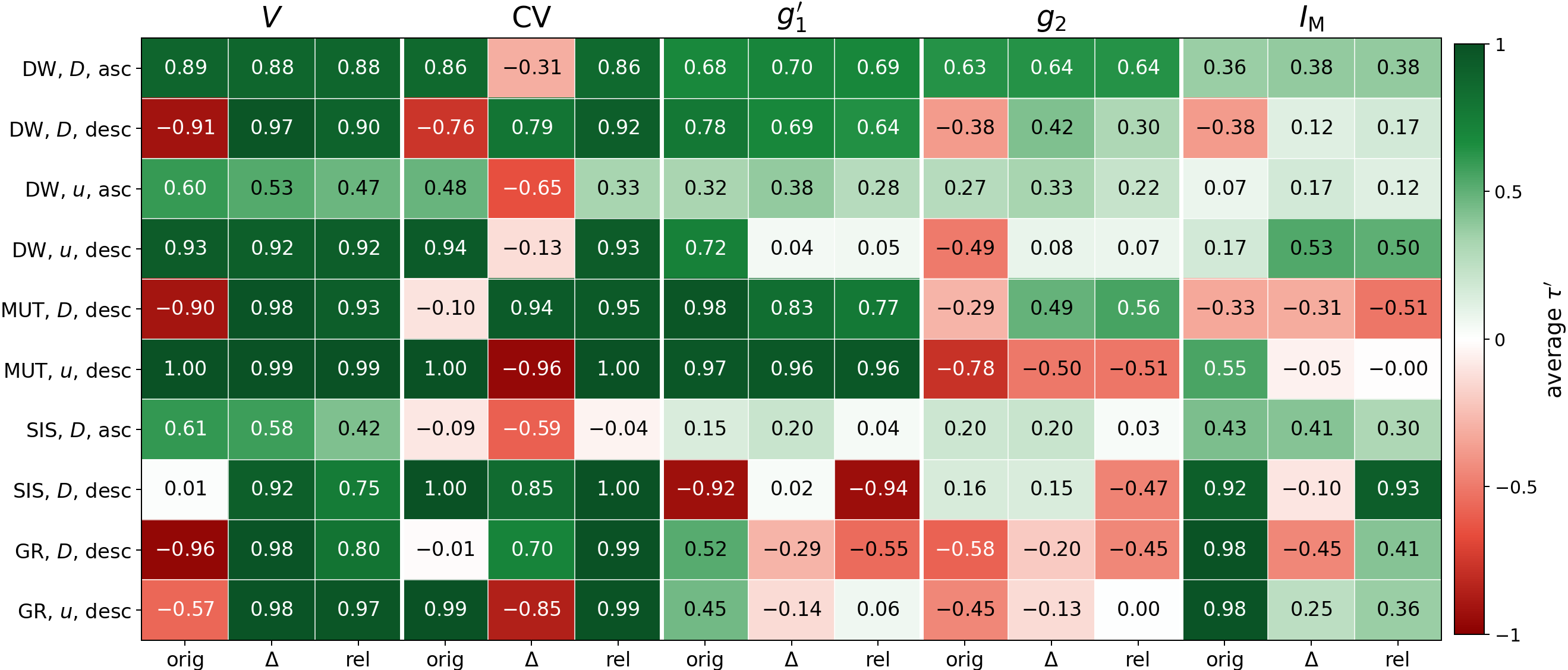}
\caption{Average sign-adjusted Kendall's $\tau$ (i.e., $\tau'$) over the 40 networks, computed over the second half of the home range. The rows correspond to the ten simulation conditions (DW: double-well; MUT: mutualistic interaction; SIS: susceptible--infectious--susceptible; GR: gene-regulatory; $D$ or $u$: control parameter; asc/desc: ascending/descending). Each column corresponds to one of the five classical spatial EWSs in one of three variants: original (orig), difference ($\Delta$, computed on $x_i^*-b_i$), or ratio (rel, computed on $x_i^*/b_i$).}
\label{fig:si-half}
\end{figure}

In Fig.~3 of the main text, we evaluated the five classical spatial EWSs and their two baseline-referenced variants using the sign-adjusted Kendall's $\tau'$, computed over the full home range. Here, we confirm that the conclusions are unchanged if we instead compute $\tau'$ over the second half of the home range, i.e., the half closer to the tipping point. Figure~\ref{fig:si-half} shows the average $\tau'$ over the 40 networks for the ten simulation conditions and the fifteen EWS--variant combinations, exactly as in the main text except that $\tau'$ is computed only over the second half of the home range. The pattern is qualitatively the same as that for the full home range. In particular, the original spatial variance and CV are unreliable, whereas $V_{\Delta}$ and $V_{\rm rel}$ are positive in all ten simulation conditions and $\text{CV}_{\rm rel}$ is positive in nine of them. The single exception is $\text{CV}_{\rm rel}$ for the ascending SIS dynamics, for which $\tau'$ is slightly negative over the second half of the home range. This simulation condition is the same as that in which $\text{CV}_{\rm rel}$ is already the weakest EWS over the full home range.

\clearpage

\section{Comparison with single-node temporal EWSs}\label{sec:si-temporal}

In the main text, we compare our spatial EWSs with a classical single-node temporal EWS as a function of the number of temporal samples, $L'$, using the descending-$D$ simulation for the coupled double-well dynamics as an example (Fig.~5 of the main text). Here, we provide the same comparison for all ten simulation conditions.

\begin{figure}[htbp]\centering
\includegraphics[width=0.9\textwidth]{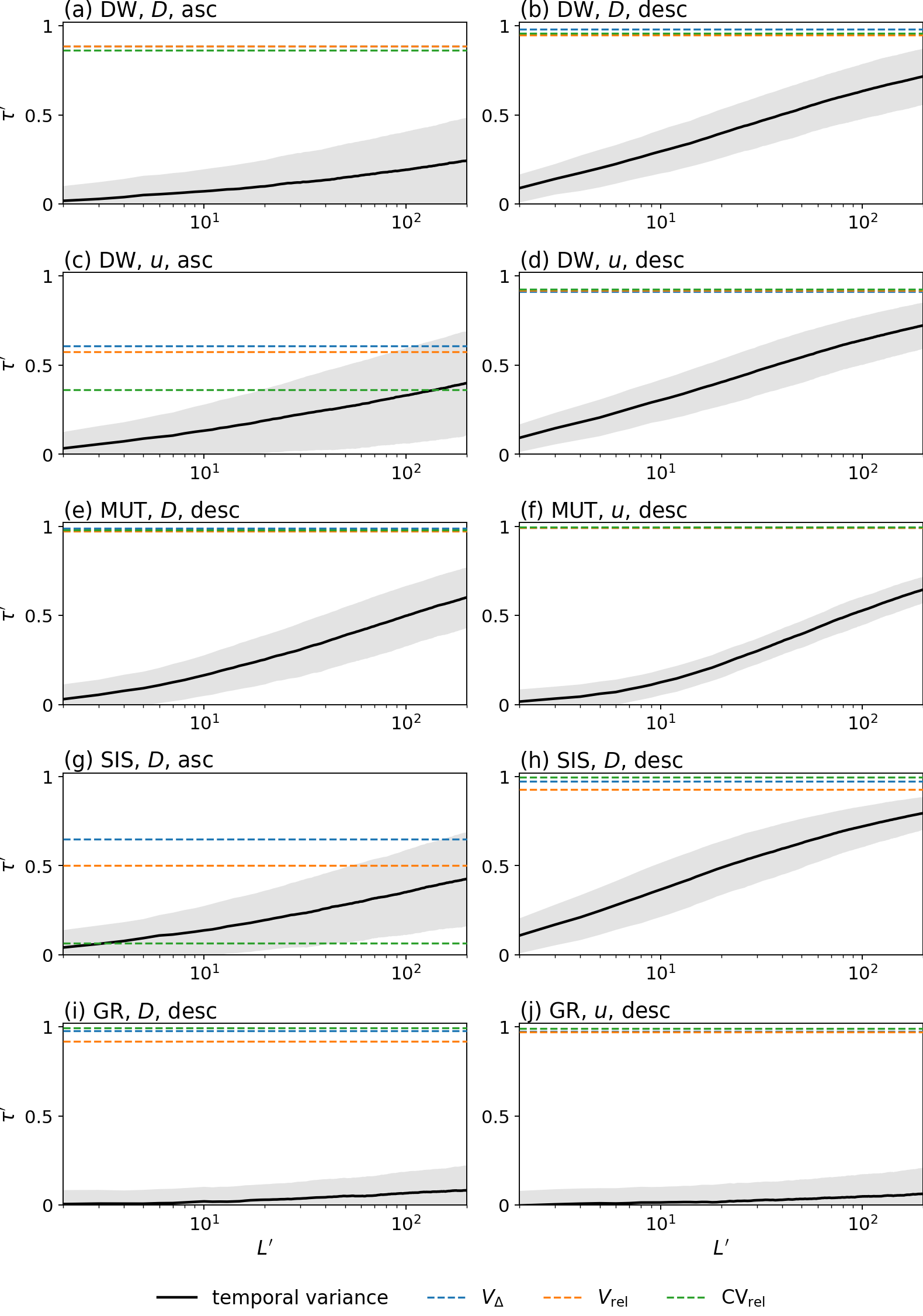}
\caption{Sign-adjusted Kendall's $\tau$ (i.e., $\tau'$) compared between the single-node temporal variance and the three baseline-referenced spatial EWSs. Each panel corresponds to one simulation condition. Panel~(b) is the same as Fig.~5(a) of the main text. The black solid curve shows the average over the 40 networks and the ten randomly chosen nodes per network, i.e., over $400$ network--node pairs. The shaded area shows one standard deviation above and below the average, also computed over the $400$ network--node pairs. The dashed horizontal lines indicate $V_\Delta$ (blue), $V_{\rm rel}$ (orange), and $\mathrm{CV}_{\rm rel}$ (green), which do not depend on the number of samples, $L'$, used to compute the temporal variance. In several panels, two or all three spatial reference lines nearly or completely overlap, so fewer than three dashed lines are visible.}
\label{fig:si-temp-tau}
\end{figure}

\begin{figure}[htbp]\centering
\includegraphics[width=0.9\textwidth]{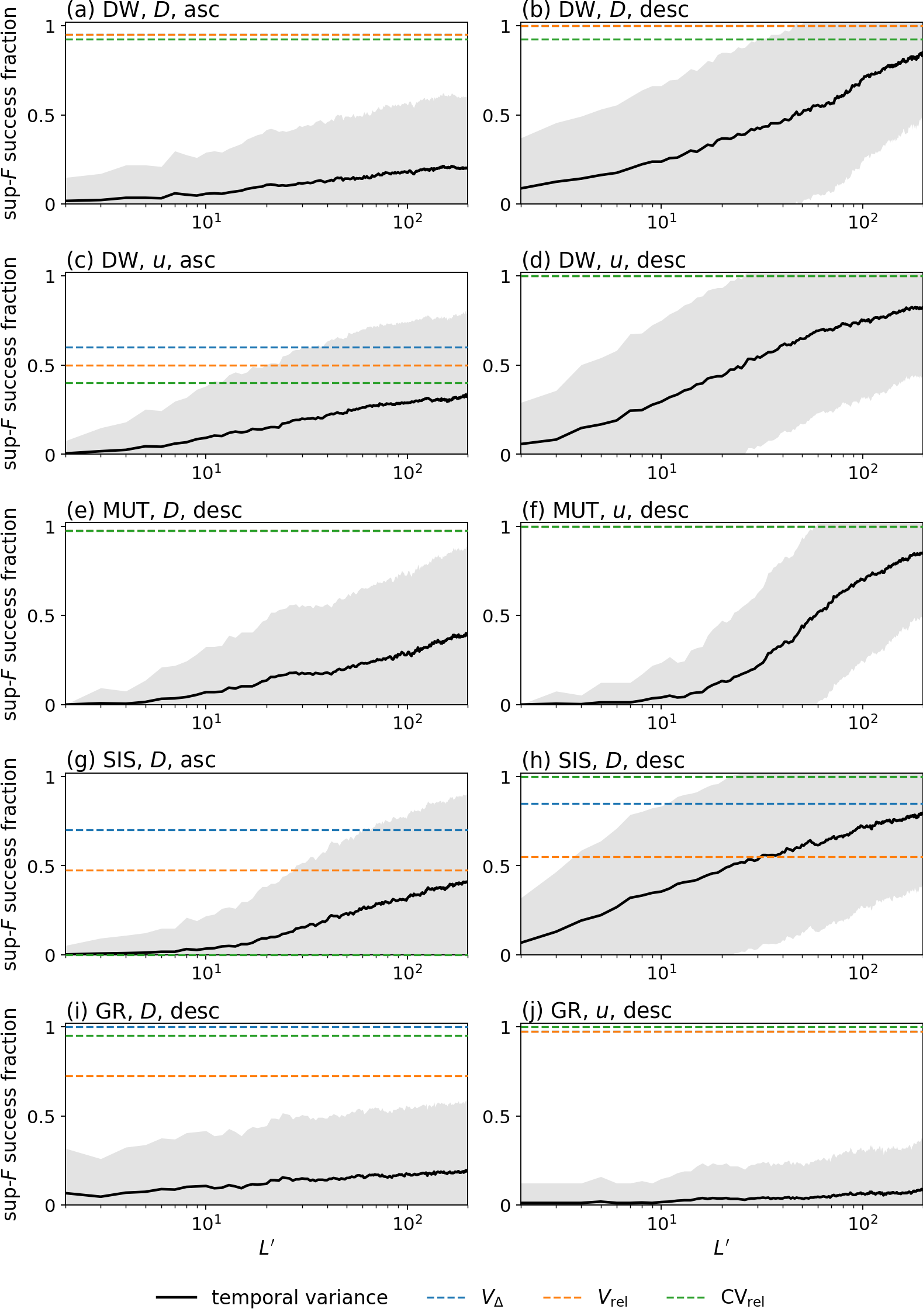}
\caption{Non-sequential sup-$F$ success fraction compared between the single-node temporal variance and the three baseline-referenced spatial EWSs. Each panel corresponds to one simulation condition. Panel~(b) is the same as Fig.~5(b) of the main text. See the caption of Fig.~\ref{fig:si-temp-tau} for the plotting conventions.}
\label{fig:si-temp-ns}
\end{figure}

\begin{figure}[htbp]\centering
\includegraphics[width=0.9\textwidth]{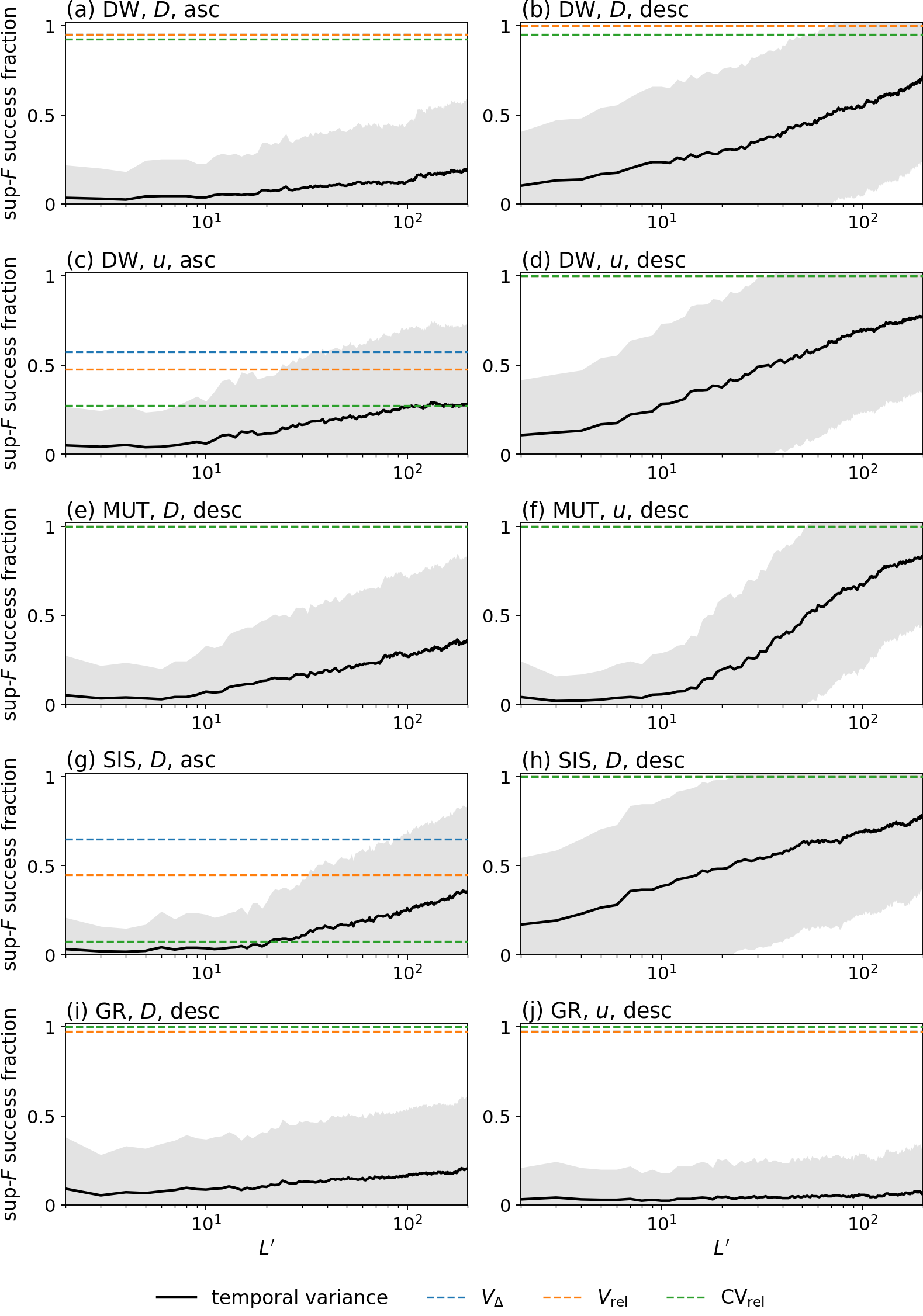}
\caption{Sequential sup-$F$ success fraction compared between the single-node temporal variance and the three baseline-referenced spatial EWSs. Each panel corresponds to one simulation condition. Panel~(b) is the same as Fig.~5(c) of the main text. See the caption of Fig.~\ref{fig:si-temp-tau} for the plotting conventions.}
\label{fig:si-temp-sq}
\end{figure}

Figures~\ref{fig:si-temp-tau}, \ref{fig:si-temp-ns}, and \ref{fig:si-temp-sq} show $\tau'$, the non-sequential sup-$F$ success fraction, and the sequential sup-$F$ success fraction, respectively, as the performance measures. We compare the temporal variance with the three baseline-referenced spatial EWSs, $V_\Delta$, $V_{\rm rel}$, and $\mathrm{CV}_{\rm rel}$. In each figure, one panel corresponds to one of the ten simulation conditions. In every simulation condition, the temporal variance improves as $L'$, the number of samples used to compute the temporal variance, increases. However, the temporal variance performs worse than the spatial $V_\Delta$ throughout $L' \le 200$ for all three performance measures and under all simulation conditions. In terms of $\tau'$, the temporal variance does not outperform $V_{\rm rel}$ for any $L' \le 200$ in any simulation condition. It reaches or exceeds $\mathrm{CV}_{\rm rel}$ in two of the ten simulation conditions when $L'$ is larger than a condition-dependent threshold. The results are similar for the sup-$F$ success fractions. The temporal variance outperforms $V_{\rm rel}$ or $\mathrm{CV}_{\rm rel}$ in at most one or two, respectively, of the ten simulation conditions for each of the two sup-$F$-based performance measures, as shown in Figs.~\ref{fig:si-temp-ns} and \ref{fig:si-temp-sq}.

\newpage

\clearpage

\clearpage

\section{Parameters used in numerical simulations}

The parameters and settings for all simulations are provided in \texttt{simulation\_parameters.csv}. Each row corresponds to one combination of the following variables, which are listed in the columns of the file:
\begin{itemize}

\item dynamical system, i.e., coupled double-well, mutualistic species, SIS, or gene-regulatory dynamics;

\item control parameter, i.e., $D$ or $u$;

\item fixed value of the non-control parameter, i.e., ``$D$'' or ``$u$'' column. In the descending-$D$ simulations of the coupled double-well and mutualistic species dynamics, the fixed stress is $u=-5$, which is necessary to destabilize the upper state so that a downward transition occurs;

\item simulation direction, i.e., ascending or descending;

\item network, given by both its display name and the code name used in the data files, and its number of nodes $N$;

\item simulation range of the control parameter, given by ``far'', i.e., the value farthest from the bifurcation at which all nodes remain in their initial state, and ``near'', i.e., a value close to or beyond the first transition. We sample $100$ evenly spaced control-parameter values across this range;

\item home range, given by the ``home\_boundary'', i.e., the last control-parameter value before the first node tips, and the ``home\_len'', i.e., the number of control-parameter values that lie within the home range. The EWSs are computed only within the home range; and

\item $\Delta t$, the time step for numerical integration.

\end{itemize}

The dynamical-system models, all parameter values, and the fixed values of the non-control parameters are identical to those in our previous study~\cite{maclaren2025applicability-SI} for every row of the table. For each simulation condition that also appears in~\cite{maclaren2025applicability-SI}, i.e., for the $36$ networks shared with that study, we reused the same simulation range and time step $\Delta t$ whenever they still produced a transition within the range. We re-determined the simulation range only when necessary, namely, for the four newly added networks and for the few shared conditions in which the previous range did not contain a transition. In these cases, we determined the simulation range anew so that it contained a transition and the home range was well defined. For the four largest new networks, we also reduced $\Delta t$ for numerical stability.

\newpage



\begin{thebibliography}{10}

\bibitem{Scheffer2001Nature}
M.~Scheffer, S.~Carpenter, J.~A. Foley, C.~Folke, and B.~Walker.
\newblock Catastrophic shifts in ecosystems.
\newblock {\em Nature}, 413:591--596, 2001.

\bibitem{Lenton2008PNAS}
T.~M. Lenton, H.~Held, E.~Kriegler, J.~W. Hall, W.~Lucht, S.~Rahmstorf, and
  H.~J. Schellnhuber.
\newblock Tipping elements in the {E}arth's climate system.
\newblock {\em Proceedings of the National Academy of Sciences of the United
  States of America}, 105:1786--1793, 2008.

\bibitem{Scheffer2009Nature}
M.~Scheffer, J.~Bascompte, W.~A. Brock, V.~Brovkin, S.~R. Carpenter, V.~Dakos,
  H.~Held, E.~H. van Nes, M.~Rietkerk, and G.~Sugihara.
\newblock Early-warning signals for critical transitions.
\newblock {\em Nature}, 461:53--59, 2009.

\bibitem{dakos2012methods}
V.~Dakos, S.~R. Carpenter, W.~A. Brock, A.~M. Ellison, V.~Guttal, A.~R. Ives,
  S.~K{\'e}fi, V.~Livina, D.~A. Seekell, E.~H. van Nes, and M.~Scheffer.
\newblock Methods for detecting early warnings of critical transitions in time
  series illustrated using simulated ecological data.
\newblock {\em PLoS ONE}, 7:e41010, 2012.

\bibitem{dakos2015resilience}
V.~Dakos, S.~R. Carpenter, E.~H. van Nes, and M.~Scheffer.
\newblock Resilience indicators: prospects and limitations for early warnings
  of regime shifts.
\newblock {\em Philosophical Transactions of the Royal Society B},
  370:20130263, 2015.

\bibitem{boettiger2013early}
C.~Boettiger and A.~Hastings.
\newblock Early warning signals and the prosecutor’s fallacy.
\newblock {\em Ecology Letters}, 16:271--280, 2013.

\bibitem{kefi2013early}
S.~K{\'e}fi, V.~Dakos, M.~Scheffer, E.~H. Van~Nes, and M.~Rietkerk.
\newblock Early warning signals also precede non-catastrophic transitions.
\newblock {\em Oikos}, 122:641--648, 2013.

\bibitem{rietkerk2025ambiguity}
M.~Rietkerk, V.~Skiba, E.~Weinans, R.~H\'{e}bert, and T.~Laepple.
\newblock Ambiguity of early warning signals for climate tipping points.
\newblock {\em Nature Climate Change}, 15:479--488, 2025.

\bibitem{Dakos2012TheorEcol}
V.~Dakos, S.~R. Carpenter, W.~A. Brock, A.~M. Ellison, V.~Guttal, A.~R. Ives,
  S.~K{\'e}fi, V.~Livina, D.~A. Seekell, E.~H. van Nes, and M.~Scheffer.
\newblock Generic indicators of critical transitions: methods for detection.
\newblock {\em Theoretical Ecology}, 5:141--154, 2012.

\bibitem{Patterson2021AmNat}
A.~C. Patterson, A.~G. Strang, and K.~C. Abbott.
\newblock When and where we can expect to see early warning signals in
  multispecies systems approaching tipping points: insights from theory.
\newblock {\em American Naturalist}, 198:E12--E26, 2021.

\bibitem{Aparicio2021PNAS}
A.~Aparicio, J.~X. Velasco-Hern{\'a}ndez, C.~H. Moog, Y.-Y. Liu, and M.~T.
  Angulo.
\newblock Structure-based identification of sensor species for anticipating
  critical transitions.
\newblock {\em Proceedings of the National Academy of Sciences of the United
  States of America}, 118:e2104732118, 2021.

\bibitem{MacLaren2023JRoySocInterface}
N.~G. MacLaren, P.~Kundu, and N.~Masuda.
\newblock Early warnings for multi-stage transitions in dynamics on networks.
\newblock {\em Journal of the Royal Society Interface}, 20:20220743, 2023.

\bibitem{masuda2024anticipating}
N.~Masuda, K.~Aihara, and N.~G. MacLaren.
\newblock Anticipating regime shifts by mixing early warning signals from
  different nodes.
\newblock {\em Nature Communications}, 15:1086, 2024.

\bibitem{boettiger2012quantifying}
C.~Boettiger and A.~Hastings.
\newblock Quantifying limits to detection of early warning for critical
  transitions.
\newblock {\em Journal of the Royal Society Interface}, 9:2527--2539, 2012.

\bibitem{hessler2022bayesian}
M.~He{\ss}ler and O.~Kamps.
\newblock Bayesian on-line anticipation of critical transitions.
\newblock {\em New Journal of Physics}, 24:063021, 2022.

\bibitem{kong2021machine}
L.-W. Kong, H.-W. Fan, C.~Grebogi, and Y.-C. Lai.
\newblock Machine learning prediction of critical transition and system
  collapse.
\newblock {\em Physical Review Research}, 3:013090, 2021.

\bibitem{patel2023using}
D.~Patel and E.~Ott.
\newblock Using machine learning to anticipate tipping points and extrapolate
  to post-tipping dynamics of nonstationary dynamical systems.
\newblock {\em Chaos}, 33:023143, 2023.

\bibitem{liu2024early}
Z.~Liu, X.~Zhang, X.~Ru, T.-T. Gao, J.~M. Moore, and G.~Yan.
\newblock Early predictor for the onset of critical transitions in networked
  dynamical systems.
\newblock {\em Physical Review X}, 14:031009, 2024.

\bibitem{JCGM100_2008}
{JCGM/WG 1}.
\newblock Evaluation of measurement data --- guide to the expression of
  uncertainty in measurement (gum 1995 with minor corrections).
\newblock JCGM 100:2008, 2008.
\newblock Subclause 4.3.2 (Note) and Table E.1. Available at
  \url{https://www.bipm.org/documents/20126/2071204/JCGM_100_2008_E.pdf}.

\bibitem{Biggs2009PNAS}
R.~Biggs, S.~R. Carpenter, and W.~A. Brock.
\newblock Turning back from the brink: detecting an impending regime shift in
  time to avert it.
\newblock {\em Proceedings of the National Academy of Sciences of the United
  States of America}, 106:826--831, 2009.

\bibitem{Dakos2010TheorEcol}
V.~Dakos, E.~H. van Nes, R.~Donangelo, H.~Fort, and M.~Scheffer.
\newblock Spatial correlation as leading indicator of catastrophic shifts.
\newblock {\em Theoretical Ecology}, 3:163--174, 2010.

\bibitem{buelo2022evaluating}
C.~D. Buelo, M.~L. Pace, S.~R. Carpenter, E.~H. Stanley, D.~A. Ortiz, and D.~T.
  Ha.
\newblock Evaluating the performance of temporal and spatial early warning
  statistics of algal blooms.
\newblock {\em Ecological Applications}, 32:e2616, 2022.

\bibitem{gsell2016evaluating}
A.~S. Gsell, U.~Scharfenberger, D.~{\"O}zkundakci, A.~Walters, L.-A. Hansson,
  A.~B.~G. Janssen, P.~N{\~o}ges, P.~C. Reid, D.~E. Schindler, E.~Van~Donk,
  V.~Dakos, and R.~Adrian.
\newblock Evaluating early-warning indicators of critical transitions in
  natural aquatic ecosystems.
\newblock {\em Proceedings of the National Academy of Sciences of the United
  States of America}, 113:E8089--E8095, 2016.

\bibitem{curtiss2023rising}
J.~E. Curtiss, D.~Mischoulon, L.~B. Fisher, C.~Cusin, S.~Fedor, R.~W. Picard,
  and P.~Pedrelli.
\newblock Rising early warning signals in affect associated with future changes
  in depression: a dynamical systems approach.
\newblock {\em Psychological Medicine}, 53:3124--3132, 2023.

\bibitem{dai2013slower}
L.~Dai, K.~S. Korolev, and J.~Gore.
\newblock Slower recovery in space before collapse of connected populations.
\newblock {\em Nature}, 496:355--358, 2013.

\bibitem{kefi2014early}
S.~K{\'e}fi, V.~Guttal, W.~A. Brock, S.~R. Carpenter, A.~M. Ellison, V.~N.
  Livina, D.~A. Seekell, M.~Scheffer, E.~H. van Nes, and V.~Dakos.
\newblock Early warning signals of ecological transitions: methods for spatial
  patterns.
\newblock {\em PLoS ONE}, 9:e92097, 2014.

\bibitem{nijp2019spatial}
J.~J. Nijp, A.~J. A.~M. Temme, G.~A.~K. van Voorn, L.~Kooistra, G.~M.
  Hengeveld, M.~B. Soons, A.~J. Teuling, and J.~Wallinga.
\newblock Spatial early warning signals for impending regime shifts: a
  practical framework for application in real-world landscapes.
\newblock {\em Global Change Biology}, 25:1905--1921, 2019.

\bibitem{fernandez2009catastrophic}
A.~Fern{\'a}ndez and H.~Fort.
\newblock Catastrophic phase transitions and early warnings in a spatial
  ecological model.
\newblock {\em Journal of Statistical Mechanics}, 2009:P09014, 2009.

\bibitem{storch2022topological}
L.~S. Storch and S.~L. Day.
\newblock Topological early warning signals: quantifying varying routes to
  extinction in a spatially distributed population model.
\newblock {\em Journal of Theoretical Biology}, 554:111274, 2022.

\bibitem{Newman2018book}
M.~E.~J. Newman.
\newblock {\em Networks}.
\newblock Oxford University Press, Oxford, UK, second edition, 2018.

\bibitem{maclaren2025applicability}
N.~G. MacLaren, K.~Aihara, and N.~Masuda.
\newblock Applicability of spatial early warning signals to complex network
  dynamics.
\newblock {\em Journal of the Royal Society Interface}, 22:20240696, 2025.

\bibitem{robinson2025assessing}
G.~E. Robinson and G.~M. Donovan.
\newblock Assessing and comparing early warning signal performance in
  spatially-structured systems.
\newblock {\em PLoS ONE}, 20:e0332695, 2025.

\bibitem{brummitt2015coupled}
C.~D. Brummitt, G.~Barnett, and R.~M. D'Souza.
\newblock Coupled catastrophes: sudden shifts cascade and hop among
  interdependent systems.
\newblock {\em Journal of the Royal Society Interface}, 12:20150712, 2015.

\bibitem{wunderling2022recurrent}
N.~Wunderling, A.~Staal, B.~Sakschewski, M.~Hirota, O.~A. Tuinenburg, J.~F.
  Donges, H.~M.~J. Barbosa, and R.~Winkelmann.
\newblock Recurrent droughts increase risk of cascading tipping events by
  outpacing adaptive capacities in the {A}mazon rainforest.
\newblock {\em Proceedings of the National Academy of Sciences of the United
  States of America}, 119:e2120777119, 2022.

\bibitem{guttal2009spatial}
V.~Guttal and C.~Jayaprakash.
\newblock Spatial variance and spatial skewness: leading indicators of regime
  shifts in spatial ecological systems.
\newblock {\em Theoretical Ecology}, 2:3--12, 2009.

\bibitem{buelo2018modeling}
C.~D. Buelo, S.~R. Carpenter, and M.~L. Pace.
\newblock A modeling analysis of spatial statistical indicators of thresholds
  for algal blooms.
\newblock {\em Limnology and Oceanography Letters}, 3:384--392, 2018.

\bibitem{eby2017alternative}
S.~Eby, A.~Agrawal, S.~Majumder, A.~P. Dobson, and V.~Guttal.
\newblock Alternative stable states and spatial indicators of critical slowing
  down along a spatial gradient in a savanna ecosystem.
\newblock {\em Global Ecology and Biogeography}, 26:638--649, 2017.

\bibitem{litzow2017indications}
M.~A. Litzow.
\newblock Indications of hysteresis and early warning signals of reduced
  community resilience during a {B}ering {S}ea cold anomaly.
\newblock {\em Marine Ecology Progress Series}, 571:13--28, 2017.

\bibitem{litzow2008increased}
M.~A. Litzow, J.~D. Urban, and B.~J. Laurel.
\newblock Increased spatial variance accompanies reorganization of two
  continental shelf ecosystems.
\newblock {\em Ecological Applications}, 18:1331--1337, 2008.

\bibitem{rindi2018experimental}
L.~Rindi, M.~Dal~Bello, and L.~Benedetti-Cecchi.
\newblock Experimental evidence of spatial signatures of approaching regime
  shifts in macroalgal canopies.
\newblock {\em Ecology}, 99:1709--1715, 2018.

\bibitem{legendre1989spatial}
P.~Legendre and M.~J. Fortin.
\newblock Spatial pattern and ecological analysis.
\newblock {\em Vegetatio}, 80:107--138, 1989.

\bibitem{okabe2015spatial}
A.~Okabe and K.~Sugihara.
\newblock {\em Spatial analysis along networks: statistical and computational
  methods}.
\newblock John Wiley \& Sons, Chichester, UK, 2012.

\bibitem{Clarke2026JPhysComplexity}
J.~Clarke, C.~Huntingford, P.~D.~L. Ritchie, and P.~Cox.
\newblock Conditions for skilful spatial and temporal tipping point early
  warning signals.
\newblock {\em Journal of Physics: Complexity}, 7:025007, 2026.

\bibitem{gao2016universal}
J.~Gao, B.~Barzel, and A.-L. Barab{\'a}si.
\newblock Universal resilience patterns in complex networks.
\newblock {\em Nature}, 530:307--312, 2016.

\bibitem{lever2020foreseeing}
J.~J. Lever, I.~A. van~de Leemput, E.~Weinans, R.~Quax, V.~Dakos, E.~H. van
  Nes, J.~Bascompte, and M.~Scheffer.
\newblock Foreseeing the future of mutualistic communities beyond collapse.
\newblock {\em Ecology Letters}, 23:2--15, 2020.

\bibitem{Bury2020JRSocInterface}
T.~M. Bury, C.~T. Bauch, and M.~Anand.
\newblock Detecting and distinguishing tipping points using spectral early
  warning signals.
\newblock {\em Journal of the Royal Society Interface}, 17:20200482, 2020.

\bibitem{masuda2026tipmoc}
N.~Masuda.
\newblock Detecting and forecasting tipping points from sample variance alone.
\newblock {\em PNAS Nexus}, 5:pgag126, 2026.

\bibitem{quandt1960tests}
R.~E. Quandt.
\newblock Tests of the hypothesis that a linear regression system obeys two
  separate regimes.
\newblock {\em Journal of the American Statistical Association}, 55:324--330,
  1960.

\bibitem{andrews1993tests}
D.~W.~K. Andrews.
\newblock Tests for parameter instability and structural change with unknown
  change point.
\newblock {\em Econometrica}, 61:821--856, 1993.

\bibitem{zeileis2002strucchange}
A.~Zeileis, F.~Leisch, K.~Hornik, and C.~Kleiber.
\newblock strucchange: An {R} package for testing for structural change in
  linear regression models.
\newblock {\em Journal of Statistical Software}, 7:1--38, 2002.

\bibitem{yu2026covariance}
S.~Yu, N.~G. MacLaren, and N.~Masuda.
\newblock Using covariance of node states to design early warning signals for
  network dynamics.
\newblock {\em Philosophical Transactions of the Royal Society A},
  384:20240486, 2026.

\bibitem{carpenter2006rising}
S.~R. Carpenter and W.~A. Brock.
\newblock Rising variance: a leading indicator of ecological transition.
\newblock {\em Ecology Letters}, 9:311--318, 2006.

\bibitem{harris2020early}
M.~J. Harris, S.~I. Hay, and J.~M. Drake.
\newblock Early warning signals of malaria resurgence in {K}ericho, {K}enya.
\newblock {\em Biology Letters}, 16:20190713, 2020.

\bibitem{Dakos2014PNAS}
V.~Dakos and J.~Bascompte.
\newblock Critical slowing down as early warning for the onset of collapse in
  mutualistic communities.
\newblock {\em Proceedings of the National Academy of Sciences of the United
  States of America}, 111:17546--17551, 2014.

\bibitem{Dakos2018EcolInd}
V.~Dakos.
\newblock Identifying best-indicator species for abrupt transitions in
  multispecies communities.
\newblock {\em Ecological Indicators}, 94:494--502, 2018.

\bibitem{Berlow2004JAnimEcol}
E.~L. Berlow, A.-M. Neutel, J.~E. Cohen, P.~C. de~Ruiter, B.~Ebenman,
  M.~Emmerson, J.~W. Fox, V.~A.~A. Jansen, J.~I. Jones, G.~D. Kokkoris, D.~O.
  Logofet, A.~J. McKane, J.~M. Montoya, and O.~Petchey.
\newblock Interaction strengths in food webs: issues and opportunities.
\newblock {\em Journal of Animal Ecology}, 73:585--598, 2004.

\bibitem{GrossFeudel2006PhysRevE}
T.~Gross and U.~Feudel.
\newblock Generalized models as a universal approach to the analysis of
  nonlinear dynamical systems.
\newblock {\em Physical Review E}, 73:016205, 2006.

\bibitem{Novak2016AnnuRevEcolEvolSyst}
M.~Novak, J.~D. Yeakel, A.~E. Noble, D.~F. Doak, M.~Emmerson, J.~A. Estes,
  U.~Jacob, M.~T. Tinker, and J.~T. Wootton.
\newblock Characterizing species interactions to understand press
  perturbations: what is the community matrix?
\newblock {\em Annual Review of Ecology, Evolution, and Systematics},
  47:409--432, 2016.

\bibitem{dablander2023anticipating}
F.~Dablander, A.~Pichler, A.~Cika, and A.~Bacilieri.
\newblock Anticipating critical transitions in psychological systems using
  early warning signals: theoretical and practical considerations.
\newblock {\em Psychological Methods}, 28:765--790, 2023.

\bibitem{speed2002tukey}
T.~P. Speed.
\newblock John {W}. {T}ukey's contributions to analysis of variance.
\newblock {\em Annals of Statistics}, 30:1649--1665, 2002.

\bibitem{cho2008variance}
E.~Cho and M.~J. Cho.
\newblock Variance of the with-replacement sample variance.
\newblock In {\em JSM Proceedings, Survey Research Methods Section}, pages
  1291--1293. American Statistical Association, Alexandria, VA, 2008.

\bibitem{vegas2012direct}
G.~Vegas-S\'{a}nchez-Ferrero, S.~Aja-Fern\'{a}ndez,
  M.~Mart\'{i}n-Fern\'{a}ndez, and C.~Palencia.
\newblock A direct calculation of moments of the sample variance.
\newblock {\em Mathematics and Computers in Simulation}, 82:790--804, 2012.

\bibitem{ditlevsen2023warning}
P.~Ditlevsen and S.~Ditlevsen.
\newblock Warning of a forthcoming collapse of the {A}tlantic meridional
  overturning circulation.
\newblock {\em Nature Communications}, 14:4254, 2023.

\bibitem{pastor2015epidemic}
R.~Pastor-Satorras, C.~Castellano, P.~Van~Mieghem, and A.~Vespignani.
\newblock Epidemic processes in complex networks.
\newblock {\em Reviews of Modern Physics}, 87:925--979, 2015.

\bibitem{chen2022practical}
S.~Chen, A.~Ghadami, and B.~I. Epureanu.
\newblock Practical guide to using {K}endall’s $\tau$ in the context of
  forecasting critical transitions.
\newblock {\em Royal Society Open Science}, 9:211346, 2022.

\bibitem{chow1960tests}
G.~C. Chow.
\newblock Tests of equality between sets of coefficients in two linear
  regressions.
\newblock {\em Econometrica}, 28:591--605, 1960.

\end{thebibliography}

\begin{thebibliography}{10}

\bibitem{Dakos2010TheorEcol-SI}
V.~Dakos, E.~H. van Nes, R.~Donangelo, H.~Fort, and M.~Scheffer.
\newblock Spatial correlation as leading indicator of catastrophic shifts.
\newblock {\em Theoretical Ecology}, 3:163--174, 2010.

\bibitem{kefi2014early-SI}
S.~K{\'e}fi, V.~Guttal, W.~A. Brock, S.~R. Carpenter, A.~M. Ellison, V.~N.
  Livina, D.~A. Seekell, M.~Scheffer, E.~H. van Nes, and V.~Dakos.
\newblock Early warning signals of ecological transitions: methods for spatial
  patterns.
\newblock {\em PLoS ONE}, 9:e92097, 2014.

\bibitem{maclaren2025applicability-SI}
N.~G. MacLaren, K.~Aihara, and N.~Masuda.
\newblock Applicability of spatial early warning signals to complex network
  dynamics.
\newblock {\em Journal of the Royal Society Interface}, 22:20240696, 2025.

\bibitem{kunegis2013konect-SI}
J.~Kunegis.
\newblock {KONECT}: the {K}oblenz {N}etwork {C}ollection.
\newblock In {\em Proceedings of the 22nd International Conference on the World
  Wide Web}, pages 1343--1350, 2013.

\bibitem{peixoto2020netzschleuder-SI}
T.~P. Peixoto.
\newblock The {N}etzschleuder network catalogue and repository.
\newblock {\em Zenodo}, 2020.
\newblock \url{https://networks.skewed.de/}, accessed 3 August 2026.

\bibitem{csardi2024igraph-SI}
G.~Cs{\'a}rdi, T.~Nepusz, V.~Traag, S.~Horv{\'a}t, F.~Zanini, D.~Noom, and
  K.~M{\"u}ller.
\newblock igraph: network analysis and visualization in {R}.
\newblock {\em {R} package version}, 2:10--5281, 2024.

\bibitem{hagberg2007exploring-SI}
A.~A. Hagberg, D.~A. Schult, and P.~J. Swart.
\newblock Exploring network structure, dynamics, and function using {NetworkX}.
\newblock In G.~Varoquaux, T.~Vaught, and J.~Millman, editors, {\em Proceedings
  of the 7th Python in Science Conference}, pages 11--15, Pasadena, CA USA,
  2008.

\bibitem{descormiers2011alliances-SI}
K.~Descormiers and C.~Morselli.
\newblock Alliances, conflicts, and contradictions in {M}ontreal’s street
  gang landscape.
\newblock {\em International Criminal Justice Review}, 21:297--314, 2011.

\bibitem{baird1989seasonal-SI}
D.~Baird and R.~E. Ulanowicz.
\newblock The seasonal dynamics of the {C}hesapeake {B}ay ecosystem.
\newblock {\em Ecological Monographs}, 59:329--364, 1989.

\bibitem{freeman1988human-SI}
L.~C. Freeman, S.~C. Freeman, and A.~G. Michaelson.
\newblock On human social intelligence.
\newblock {\em Journal of Social and Biological Structures}, 11:415--425, 1988.

\bibitem{knuth2005art-SI}
D.~E. Knuth.
\newblock {\em The Art of Computer Programming. Volume 4, Fascicle 0}.
\newblock Addison-Wesley, Upper Saddle River, NJ, 2005.

\bibitem{thompson2003impacts-SI}
R.~M. Thompson and C.~R. Townsend.
\newblock Impacts on stream food webs of native and exotic forest: an
  intercontinental comparison.
\newblock {\em Ecology}, 84:145--161, 2003.

\bibitem{lusseau2003bottlenose-SI}
D.~Lusseau, K.~Schneider, O.~J. Boisseau, P.~Haase, E.~Slooten, and S.~M.
  Dawson.
\newblock The bottlenose dolphin community of {D}oubtful {S}ound features a
  large proportion of long-lasting associations: can geographic isolation
  explain this unique trait?
\newblock {\em Behavioral Ecology and Sociobiology}, 54:396--405, 2003.

\bibitem{hayes2006connecting-SI}
B.~Hayes.
\newblock Connecting the dots.
\newblock {\em American Scientist}, 94:400--404, 2006.

\bibitem{correia2019city-SI}
R.~B. Correia, L.~P. de~Ara{\'u}jo~Kohler, M.~M. Mattos, and L.~M. Rocha.
\newblock City-wide electronic health records reveal gender and age biases in
  administration of known drug--drug interactions.
\newblock {\em npj Digital Medicine}, 2:74, 2019.

\bibitem{haraldsdottir1992preliminary-SI}
S.~Haraldsdottir, S.~Gupta, and R.~M. Anderson.
\newblock Preliminary studies of sexual networks in a male homosexual community
  in iceland.
\newblock {\em Journal of Acquired Immune Deficiency Syndromes}, 5:374--381,
  1992.

\bibitem{goh2001universal-SI}
K.-I. Goh, B.~Kahng, and D.~Kim.
\newblock Universal behavior of load distribution in scale-free networks.
\newblock {\em Physical Review Letters}, 87:278701, 2001.

\bibitem{watts1998collective-SI}
D.~J. Watts and S.~H. Strogatz.
\newblock Collective dynamics of ‘small-world’ networks.
\newblock {\em Nature}, 393:440--442, 1998.

\bibitem{albert1999emergence-SI}
A.-L. Barab\'{a}si and R.~Albert.
\newblock Emergence of scaling in random networks.
\newblock {\em Science}, 286:509--512, 1999.

\bibitem{holme2002growing-SI}
P.~Holme and B.~J. Kim.
\newblock Growing scale-free networks with tunable clustering.
\newblock {\em Physical Review E}, 65:026107, 2002.

\bibitem{hoglund2006gene-SI}
M.~H{\"o}glund, A.~Frigyesi, and F.~Mitelman.
\newblock A gene fusion network in human neoplasia.
\newblock {\em Oncogene}, 25:2674--2678, 2006.

\bibitem{newman2006finding-SI}
M.~E.~J. Newman.
\newblock Finding community structure in networks using the eigenvectors of
  matrices.
\newblock {\em Physical Review E}, 74:036104, 2006.

\bibitem{girvan2002community-SI}
M.~Girvan and M.~E.~J. Newman.
\newblock Community structure in social and biological networks.
\newblock {\em Proceedings of the National Academy of Sciences of the United
  States of America}, 99:7821--7826, 2002.

\bibitem{coleman1957diffusion-SI}
J.~Coleman, E.~Katz, and H.~Menzel.
\newblock The diffusion of an innovation among physicians.
\newblock {\em Sociometry}, 20:253--270, 1957.

\bibitem{fire2012predicting-SI}
M.~Fire, G.~Katz, Y.~Elovici, B.~Shapira, and L.~Rokach.
\newblock Predicting student exam’s scores by analyzing social network data.
\newblock {\em Lecture Notes in Computer Sciences}, 7669:584--595, 2012.

\bibitem{beuming2005pdzbase-SI}
T.~Beuming, L.~Skrabanek, M.~Y. Niv, P.~Mukherjee, and H.~Weinstein.
\newblock Pdzbase: a protein--protein interaction database for {PDZ}-domains.
\newblock {\em Bioinformatics}, 21:827--828, 2005.

\bibitem{michalski2011matching-SI}
R.~Michalski, S.~Palus, and P.~Kazienko.
\newblock Matching organizational structure and social network extracted from
  email communication.
\newblock In {\em International Conference on Business Information Systems},
  pages 197--206, 2011.

\bibitem{chami2017social-SI}
G.~F. Chami, S.~E. Ahnert, N.~B. Kabatereine, and E.~M. Tukahebwa.
\newblock Social network fragmentation and community health.
\newblock {\em Proceedings of the National Academy of Sciences of the United
  States of America}, 114:E7425--E7431, 2017.

\bibitem{gleiser2003community-SI}
P.~M. Gleiser and L.~Danon.
\newblock Community structure in jazz.
\newblock {\em Advances in Complex Systems}, 6:565--573, 2003.

\bibitem{vsubelj2011community-SI}
L.~{\v{S}}ubelj and M.~Bajec.
\newblock Community structure of complex software systems: Analysis and
  applications.
\newblock {\em Physica A}, 390:2968--2975, 2011.

\bibitem{shen2002network-SI}
S.~S. Shen-Orr, R.~Milo, S.~Mangan, and U.~Alon.
\newblock Network motifs in the transcriptional regulation network of
  \textit{{E}scherichia coli}.
\newblock {\em Nature Genetics}, 31:64--68, 2002.

\bibitem{de2014navigability-SI}
M.~De~Domenico, A.~Sol{\'e}-Ribalta, S.~G{\'o}mez, and A.~Arenas.
\newblock Navigability of interconnected networks under random failures.
\newblock {\em Proceedings of the National Academy of Sciences of the United
  States of America}, 111:8351--8356, 2014.

\bibitem{sun2016predicting-SI}
J.~Sun, J.~Kunegis, and S.~Staab.
\newblock Predicting user roles in social networks using transfer learning with
  feature transformation.
\newblock In {\em 2016 IEEE 16th International Conference on Data Mining
  Workshops (ICDMW)}, pages 128--135. IEEE, 2016.

\bibitem{isella2011s-SI}
L.~Isella, J.~Stehl{\'e}, A.~Barrat, C.~Cattuto, J.-F. Pinton, and W.~Van~den
  Broeck.
\newblock What's in a crowd? {A}nalysis of face-to-face behavioral networks.
\newblock {\em Journal of Theoretical Biology}, 271:166--180, 2011.

\bibitem{jeong2000large-SI}
H.~Jeong, B.~Tombor, R.~Albert, Z.~N. Oltvai, and A.-L. Barab{\'a}si.
\newblock The large-scale organization of metabolic networks.
\newblock {\em Nature}, 407:651--654, 2000.

\bibitem{cook2019whole-SI}
S.~J. Cook, T.~A. Jarrell, C.~A. Brittin, Y.~Wang, A.~E. Bloniarz, M.~A.
  Yakovlev, K.~C.~Q. Nguyen, L.~T.-H. Tang, E.~A. Bayer, J.~S. Duerr, H.~E.
  B{\"u}low, O.~Hobert, D.~H. Hall, and S.~W. Emmons.
\newblock Whole-animal connectomes of both \textit{{C}aenorhabditis elegans}
  sexes.
\newblock {\em Nature}, 571:63--71, 2019.

\bibitem{milo2002network-SI}
R.~Milo, S.~Shen-Orr, S.~Itzkovitz, N.~Kashtan, D.~Chklovskii, and U.~Alon.
\newblock Network motifs: simple building blocks of complex networks.
\newblock {\em Science}, 298:824--827, 2002.

\bibitem{hidalgo2007product-SI}
C.~A. Hidalgo, B.~Klinger, A.-L. Barab{\'a}si, and R.~Hausmann.
\newblock The product space conditions the development of nations.
\newblock {\em Science}, 317:482--487, 2007.

\bibitem{guimera2003self-SI}
R.~Guimer{\`a}, L.~Danon, A.~D{\'\i}az-Guilera, F.~Giralt, and A.~Arenas.
\newblock Self-similar community structure in a network of human interactions.
\newblock {\em Physical Review E}, 68:065103, 2003.

\end{thebibliography}
\end{document}